\newcommand\lsim{\mathrel{\rlap{\lower4pt\hbox{\hskip1pt$\sim$}}
    \raise1pt\hbox{$<$}}}
\newcommand\gsim{\mathrel{\rlap{\lower4pt\hbox{\hskip1pt$\sim$}}
    \raise1pt\hbox{$>$}}}

\documentclass[prd,showpacs,byrevtex]{revtex4}

\usepackage{graphics}

\def\dddot#1{\mathinner{\buildrel\vbox{\kern5pt\hbox{...}}\over{#1}}}

\def\g{\gamma}

\def\be{\begin{equation}}
\def\ee{\end{equation}}
\def\bq{\begin{eqnarray}}
\def\eq{\end{eqnarray}}
\def\beq{\begin{eqnarray*}}
\def\eeq{\end{eqnarray*}}

\def\bs{\begin{subequations}}
\def\es{\end{subequations}}
\def\ben{\begin{eqalignno}}
\def\een{\end{eqalignno}}

\def\({\left(}
\def\){\right)}

\begin{document}

\title{Fluctuations in the CMB Induced by Cosmic Strings: Methods
and Formalism}

\author{M.~Landriau}
\altaffiliation{Current address: Laboratoire de l'Acc\'{e}l\'{e}rateur
Lin\'{e}aire, IN2P3-CNRS et Universit\'{e} Paris-Sud, B.P. 34, 91898 Orsay
Cedex, France.}
\email{landriau@lal.in2p3.fr}
\author{E.P.S.~Shellard}
\email{E.P.S.Shellard@damtp.cam.ac.uk}

\affiliation{Department of Applied Mathematics and Theoretical Physics\\
Centre for Mathematical Sciences, University of Cambridge\\
Wilberforce Road, Cambridge CB3 0WA, U.K.}

\begin{abstract}
We present methods to compute maps of CMB fluctuations from high resolution
cosmic string networks using a full Boltzmann code, on both large and
small angular scales.  The accuracy and efficency of these methods are
discussed.  
\end{abstract}


\date{January 20, 2003}

\maketitle

\section{Introduction}\label{secintro}

The potential role of cosmic strings and other topological defects in
cosmology has been the subject of considerable interest for well over
two decades (for a review see ref.~\cite{vilenkin94}).  Perhaps the
most exciting prospect would be the detection of their distinct
observational signatures in the cosmic microwave sky.  Cosmic strings,
for example, are expected to create line-like discontinuities in the
CMB temperature pattern, whereas other defects such as global
monopoles or textures create `hot spots'.  Their discovery would
provide unprecedented information about the nature of unification in
the early universe, while their absence from the CMB would
significantly strengthen constraints on a wide range of models.

To many, the publication of the BOOMERanG
results~\cite{bernardis2000}, in particular, signaled the demise of
topological defects in cosmology.  Indeed, the detection of an
acoustic peak around $\ell\simeq 200$ was seen as evidence that
primordial adiabatic perturbations were the seeds for large-scale
structure formation, a view that has been strengthened with the
apparent resolution of further peaks (see
also~\cite{hanany2000,scott2002,pearson2002}).  However, the presence
of defects is not incompatible with inflation and post-BOOMERanG
analyses, such as refs.~\cite{bouchet2002,contaldi2000}, concluded
that they could not be ruled out. Current data allows defects to play
a significant (but subdominant) role in large-scale structure
formation.  In this sense, it is of great importance to accurately
characterize nonGaussian signals from strings, as they are likely to
provide the only direct method of detection.

In this paper we will detail the methods that we have developed to
create full-sky and high resolution CMB maps generated by cosmic
defects or any other `causal' or `active' sources.  First, in
\S\ref{sec_pert} we detail
the large set of perturbation equations that have to be solved,
following this in \S\ref{sec-sources} with a discussion of the
treatment of the source terms which distinguish this analysis from
that for inflationary fluctuations. In \S\ref{sec-numerics} we then
discuss efficient numerical implementation of CMB map-making using the
analogue of Green's function techniques, without which the problem
would not be tractable computationally.  However, we complete this
introduction by discussing previous work on cosmic defects and the
CMB, pointing out its relationship to this paper.

Some of the earliest work featured analytic results obtained for
simple string configurations \cite{kaiser84b,gott85,veeraraghavan92}.
Such exact solutions are important for testing computational methods.
However, although these analytic results are interesting, numerical
simulations are essential to obtain accurate quantitative predictions
in more general contexts.  The main drawback of numerical results in
this context is their limited dynamic range, restricted by the
light-crossing times of cosmic defect simulations.  All-sky (large
angle) CMB maps can be generated and have been used to obtain the
normalization of the power spectrum to COBE, but their angular
resolution has been poor.  On the other hand, small angle maps permit
the very important characterization of nonGaussian signals due to
defects.  Given the prospect of high resolution all-sky observations
from the MAP and Planck satellites, ideally one would aim to compute
all-sky defect maps of corresponding resolution, but computational
resources remain insufficient for this task at present.

Probably the earliest attempt at computing realistic CMB patterns
generated by defects was that of Bouchet, Bennett \&
Stebbins~\cite{bouchet88}.  They employed a flat-space formalism to
calculate the CMB temperature $\Delta T/T$ in the direction $\hat{\bf
n}$, solving the metric perturbation equations using Green's functions
$G({\bf k},t,t')$ schematically as:
\begin{equation}
\frac{\Delta T}{T}(\mathbf{\hat{n}},{\bf k},t) \propto \int
G_{\mu\nu}(\hat{\bf n},{\bf k},t,t^{\prime})\Theta_{\mu\nu}({\bf k},t')
dt^{\prime}\, ,
\end{equation}
where $\Theta_{\mu\nu}$ is the energy-momentum tensor of cosmic
strings.  Their methods neglected many effects, notably the presence
of baryons and the expansion of the Universe, concentrating solely on
the integrated Sachs-Wolfe effect.  Pen, Spergel \& Turok~\cite{pen94}
computed all-sky maps (at COBE resolution) produced by different
global defects including an approximate treatment of CDM, baryons and
radiation and their work was extended on intermediate angular scales
in ref.~\cite{coulson94}.  COBE resolution maps generated by local
cosmic strings were presented in~\cite{allen96} using the
Allen-Shellard (AS) string code~\cite{allen90}.  The power spectrum
for this map was evaluated for $\ell\leq 20$ (using an ensemble of 192
realisations) and they inferred the string linear energy density to be
$G\mu/c^2 = 1.05^{+0.35}_{-0.20}\times 10^{-6}$.

The most recent work on CMB fluctuations in the presence of causal
seeds have made use of full Boltzmann codes (see
section~\ref{sec_pert}), thus including all the relevant physics (to
first order).  The AS string code was employed again in the full
Boltzmann analysis in ref.~\cite{allen97}, in which power spectra were
computed from the brightness distribution, thus bypassing the maps.
Power spectra were computed from simulations of different cosmological
epochs and provided clear evidence of the importance of vector and
tensor modes in these models, as well as the apparent absence of
strongly defined acoustic peaks.

An alternative line-of-sight approach was used in ref.~\cite{pen97},
and also later in refs.~\cite{durrer99,contaldi99}, to calculate power
spectra for global defects.  Here, the idea was to use unequal time
correlators (UETCs) of the defect energy-momentum tensor (approximated
by an expansion in eigenvectors) as the source for the perturbation
power spectra.  In principle, the method greatly extends the available
dynamic range by exploiting the scalability of the correlators during
defect evolution.  However, while scalability is approached
asymptotically in the radiation and matter eras, during the important
radiation-matter transition the UETCs must still be calculated from
large simulations bridging this time period.  The line-of-sight method
has also been used by~\cite{albrecht99,pogosian99} who employed an
ensemble of toy model realisations of a string network and averaged
the power spectra.  The line-of-sight method can be used to compute
maps as well: Simatos
\& Perivaropoulos~\cite{simatos2001} modified it using
a more general expansion of plane waves to accommodate for phase
differences in a toy model for wiggly strings.  However, while the
method is phenomenologically interesting it was necessary to make a
number of assumptions about the string perturbation phases.

It is important to note that none of these methods is perfect, and in
some sense, they are complimentary: The direct approach developed
further here, solving the full Boltzmann equations on a
three-dimensional grid, provides reliable high resolution CMB maps.
However, the UETC method with a greater dynamic range provides a more
extensive view of the angular power spectrum.

\section{Cosmological Perturbations} \label{sec_pert}

In this section, we shall derive the equations that describe the
evolution of first order perturbations in the metric and the
energy-momentum tensor of matter fields in the presence of causal
seeds such as cosmic strings.  The formalism used is similar to that
of~\cite{pen94}, except here, the full Boltzmann equation for
relativistic matter (photons, neutrinos) is used. The treatment of the
Boltzmann distributions presented here follows on from the approach
used for scalar modes by~\cite{ma95}, but is extended to vector and
tensor modes.  The treatment of photon polarization follows the
approach used by~\cite{melchiorri96} for scalar and tensor modes, but
is extended to vector modes.  Similar, though not as general, methods
were used by~\cite{durrer99}, where scalar and vector modes were
treated in a gauge invariant formalism.

\subsection{Scalar - Vector - Tensor Decomposition}\label{sec_SVT}

In this paper, we will restrict ourselves to flat FRW models because
present cosmic defect simulations are restricted to these.  However,
this is a significant simplification which enables us to expand all
perturbations in terms of Fourier modes.  In Fourier space, a tensor
quantity $T_{ij}$ can be decomposed into scalar $T,T^S$, vector
$T^V_{i}$ and tensor components $T^T_{ij}$ in the following way:
\begin{equation}
T_{ij}(\mathbf{k}) = \frac{1}{3}\delta_{ij}T +(\hat{k_{i}}\hat{k_{i}}
-\frac{1}{3}\delta_{ij})T^{S} + (\hat{k_{i}}T_{j}^{V}
+\hat{k_{j}}T_{i}^{V}) + T^{T}_{ij} \, .
\end{equation}
This is analagous to the manner in which vector quantities can 
be decomposed into scalar and vector
components: 
\begin{equation}
V_i(\mathbf{k}) = \hat{k_{i}}V^S + V^V_i \, .
\end{equation}
In the above, vector and tensor components are tranverse
that is, $\hat{k_{i}}V^V_i=\hat{k_{i}}T^V_i=
\hat{k_{i}}\hat{k_{j}}T_{ij}^T=0$ and tensor components are,
in addition, traceless $T^T_ii= 0$.
It is also useful to express vector components in an orthonormal basis
$\mathbf{e_1}$,~$\mathbf{e_2}$ with
$\mathbf{e_1} \times \mathbf{e_2}= \mathbf{\hat{k}}$, so that
\begin{equation}
\mathbf{V}=V^S \mathbf{\hat{k}}+V^V_1 \mathbf{e_1}+V^V_2
\mathbf{e_2} \, . 
\end{equation}
We can also construct a basis for the tensor
components out of $\mathbf{e_1}$ and $\mathbf{e_2}$ by defining 
the following two matrices:
\begin{equation}\begin{array}{l}
M_+ =\mathbf{e_1}\otimes \mathbf{e_1}-
\mathbf{e_2}\otimes \mathbf{e_2}\\ 
M_\times = \mathbf{e_1}\otimes \mathbf{e_2}+
\mathbf{e_2}\otimes \mathbf{e_1} \, .
\end{array}\end{equation}
Tensor components can then be written\begin{equation}
T^T_{ij}=T^T_+ (M_+)_{ij} + T^T_\times (M_\times)_{ij} \, .
\end{equation}

\subsection{Einstein Equations}\label{sec_einstein}

We define perturbations of the conformally flat FRW metric as:
\begin{equation}
g_{\mu\nu}=a^2(\eta)(\eta_{\mu\nu}+h_{\mu\nu}) \, , \label{eq_hdef}
\end{equation}
where $\eta_{\mu\nu}$ is the Minkowski metric and $|h_{\mu\nu}|\ll 1$.  
We will work in the synchronous gauge in which
$h_{00}=h_{0i}=0$.  This gauge is not completely specified.  This will
result in extra ``constraint equations'' from the Eintein equations, to
ensure that all degrees of freedom are specified.
Perturbations in the energy-momentum tensor are given
by:
\begin{equation}\label{eq_deltat}\begin{array}{l}
\delta T^0_0 =-\delta\rho +\Theta^0_0 \\
\delta T^0_i=(\rho +p)v_i +\Theta^0_i \\
\delta T^i_j=\delta p\delta^i_j + p\Sigma^i_j +\Theta^i_j \, ,
\end{array}\end{equation}
where $\Theta^{\mu}_{\nu}$ is the energy momentum tensor of the causal
stiff source and $\Sigma^{\mu}_{\nu}$ contains the anisotropic
stresses of relativistic matter.
For the velocity terms, $v_i$ and $\Theta_{0i}$, we use slightly different
variables to those defined in section~\ref{sec_SVT}:
\begin{equation}\begin{array}{cc}
\theta = \dot{\imath}kv^S\,, & ~~~~\Theta_D = \dot{\imath}k{\mathcal{P}}^S \\
v_i^V = \dot{\imath}\tilde{v}_i^V\,, &~~~~
{\mathcal{P}}_i^V = \dot{\imath}\tilde{\mathcal{P}}_i^V \, .
\end{array}\end{equation}
These new variables enable us to write the evolution equation without
explicit $\dot{\imath}$'s, which is easier to implement numerically,
as will be discussed in section~\ref{sec-numerics}.

The equations for $h_{\mu\nu}$ are then obtained by substituting
equations (\ref{eq_hdef}) and (\ref{eq_deltat}) into the Einstein 
equations,
using the homogeneous part for the background metric.
The $00$ and $ij$ components of the equation
\begin{equation}
\delta (R_{\mu\nu}+\Lambda g_{\mu\nu})=8\pi
G\delta(T_{\mu\nu}-\frac{1}{2}g_{\mu\nu}T^{\lambda}_{\lambda})
\end{equation}
then lead, to first order,
after transforming to Fourier space and decomposing into
SVT components, to the following equations of motion for $h$ :
\begin{equation}\begin{array}{c}
\ddot{h} +\frac{\dot{a}}{a}\dot{h} = -8\pi G(a^{2}(\delta\rho +
3\delta p) +\Theta_{00} + \Theta)\,,\\
\ddot{h}^{S} + 2\frac{\dot{a}}{a}\dot{h}^{S} +\frac{k^{2}}{3}h^{-}=16\pi
G(a^{2}p\Sigma^{S} + \Theta^{S})\,,\\
\ddot{h}_{i}^{V} + 2\frac{\dot{a}}{a}\dot{h}_{i}^{V} = 16\pi
G(a^{2}p\Sigma_{i}^{V}+ \Theta_{i}^{V})\,,\\
\ddot{h}_{\epsilon}^{T} + 2\frac{\dot{a}}{a}\dot{h}_{\epsilon}^{T}
+k^{2}h_{\epsilon}^{T} 
= 16\pi G(a^{2}p\Sigma_{\epsilon}^{T}+ \Theta_{\epsilon}^{T}) \, ,
\end{array}\end{equation}
and the $00$ and $0i$ components of the equation
\begin{equation}
\delta (R_{\mu\nu}-\frac{1}{2}Rg_{\mu\nu}-\Lambda g_{\mu\nu})= 
8\pi G\delta T_{\mu\nu} 
\end{equation}
lead to the following constraint equations for $h$:
\begin{equation}\begin{array}{c}
k^{2}h^{-} + 3\frac{\dot{a}}{a}\dot{h} = 24\pi G (a^{2}\delta\rho
+\Theta_{00})\,,\\
k^2\dot{h}^{-} = 24\pi G(\Theta_D-a^{2}(\rho +p)\theta)\,,\\
k\dot{h}^V_i = 16\pi G(\tilde{\mathcal{P}}^V_i
-a^{2}(\rho+p)\tilde{v}^V_i) \, ,
\end{array}\end{equation}
where $i=1$,~$2$ and $\epsilon=+$,~$\times$ and we have defined
$h^{-}=h-h^S$.

Energy conservation ${T^{\mu\nu}}_{;\mu} = 0$ leads to the following
equations of motion for matter and radiation perturbations:
\begin{equation}\label{t_cons}\begin{array}{c}
\dot{\delta}=-(1+w)(\theta+\frac{1}{2}\dot{h})-3\frac{\dot{a}}{a}(c^2_s
-w)\delta \,,\\
\dot{\theta}=-\frac{\dot{a}}{a}(1-3c^2_s)\theta +\frac{c^2_s}{1+w}k^2
\delta +\frac{2}{3}\frac{w}{1+w}k^2 \Sigma^S \,,\\
\dot{\tilde{v}}^V_i =-\frac{\dot{a}}{a}(1-3c^2_s)\tilde{v}^V_i
-\frac{w}{1+w}k\Sigma^V_i \, ,
\end{array}\end{equation}
where $w=p/\rho$, $c^2_s=\delta p/\delta\rho$ is the sound speed
squared and $\delta=\delta\rho/\rho$.
These equations are
valid for uncoupled fluids.  And covariant energy
conservation with respect to the background 
$\Theta^{\mu\nu}_{|\mu}$ leads to
\begin{equation}\begin{array}{c}
\dot{\Theta}_{00}=-\frac{\dot{a}}{a}(\Theta_{00}+\Theta)
+\Theta_D\\
\dot{\Theta_D}=-2\frac{\dot{a}}{a}\Theta_D
-\frac{k^2}{3}(\Theta +2\Theta^S)\\
\dot{\tilde{\mathcal{P}}}^V_i=-2\frac{\dot{a}}{a}\tilde{\mathcal{P}}^V_i
+k\Theta^V_i \, .
\end{array}\end{equation}
In what follows, we shall drop the tilde on the new variables to
render the equations more legible.

\subsection{Relativistic Matter}

To treat photon perturbations, we derive the equations of motion for
the Stokes parameters, which describe polarized
light: the intensity $I$, the orientation of the polarization ellipse
$Q$ and $U$ and the ratio of its principal axis $V$.  
It is convenient to
work with the perturbations normalized to the average intensity
$I_0= \rho_\gamma/4\pi$:
\begin{equation}\begin{array}{c}
I = I_0 (1+\Delta_I) \,,\\
P = I_0\Delta_P \, ,
\end{array}\end{equation}
where $P$ stands for $Q$, $U$ or $V$.
We start with the general transfer equations for polarized light:
\begin{equation}\begin{array}{c}
\dot{\Delta}_I = + \dot{\imath}k\mu\Delta_I +
2\dot{h}_{ij}\hat{n}_i \hat{n}_j = \dot{\tau}(\Delta_I^{\mathcal{S}} -
\Delta_I + 4\mathbf{\hat{n}\cdot v_b}) \,,\\
\dot{\mathbf{\Delta}}_P + \dot{\imath}k\mu\mathbf{\Delta}_P =
\dot{\tau}(\mathbf{\Delta}_P^{\mathcal{S}} - \mathbf{\Delta}_P) \, ,
\end{array}\end{equation}
where $\mu=\mathbf{\hat{n}\cdot\hat{k}}$, $\dot{\tau}=a\sigma_T n_e$
is the differential Thompson
cross-section, $\mathbf{v_b}$ is the baryon velocity and the $\mathcal{S}$
denotes scattered quantities as measured in the comoving frame  and
the vector $\mathbf{\Delta}_P$ has $\Delta_Q$, $\Delta_U$ and
$\Delta_V$ as components.
>From the start, it is convenient to notice that $V$ is always
uncoupled to the other parameters as it cannot be generated through
Thompson scattering.  So we can set it to zero without
loss of generality.  We shall split the perturbations into scalar,
vector and tensor parts: $\Delta_X=\Delta_X^{S}
+\Delta_X^{V}+\Delta_X^{T}$, where $X=I$, $Q$ or $U$.
The treatment for massless neutrinos is identical to that for the
intensity of photons, except that all terms involving $\dot{\tau}$
are zero.

\subsubsection{Scalar perturbations}

For scalar modes, $U$ is also uncoupled from $I$ and $Q$, and hence
can be set to zero.  The contribution from the metric is
\begin{equation}
\hat{n}_i \hat{n}_j\dot{h}_{ij}^{scalar}=\frac{\dot{h}}{3}+(\mu^2
-\frac{1}{3})\dot{h}^S
\end{equation}
and the ``polarization term'' is given by
\begin{equation}
\mathbf{\Delta}_P^{S\mathcal{S}}(\mu)=
\frac{3}{16}\int_{-1}^{1} M^S(\mu,\mu^{\prime})
\mathbf{\Delta}_P^{S}(\mu^\prime) d\mu^{\prime} \, ,
\end{equation}
where the relevant block of the ``scattering matrix''
$M^S(\mu,\mu^{\prime})$ is given by
\begin{equation}
\left(\begin{array}{cc}3-\mu^{\prime 2}
-\mu^2 +3\mu^2\mu^{\prime 2}&1-\mu^{\prime 2}-3\mu^2
+3\mu^2 \mu^{\prime 2}\\1-3\mu^{\prime 2}-\mu^2 +3\mu^2\mu^{\prime
2}& 3-3\mu^{\prime 2} -3\mu^2 +3\mu^2 \mu^{\prime
2}\end{array}\right) \, ,
\end{equation}
To integrate out the angle dependence, we expand the perturbations in
Legendre polynomials:
\begin{equation}\label{legendre_exp}
\Delta^{S}_X = \sum_{\ell}(-\dot{\imath})^{\ell} (2\ell +1)
\Delta^{S}_{X\ell}P_{\ell}(\mu) \, .
\end{equation}
The equations then read
\begin{equation}\label{trans_s}\begin{array}{l}
\lefteqn{\dot{\Delta}^{S}_{I}+\dot{\imath}k\mu +\frac{2}{3}(\dot{h}+(3\mu^2
-1)\dot{h}^S) } \\
~~~~~~=\dot{\tau}(\Delta^{S}_{I0} +4\mu v_b^S
-\Delta^{S}_{I} 
-\frac{1}{2}P_2(\mu)\Pi^{S})\,, \\
\dot{\Delta}^{S}_{Q}+\dot{\imath}k\mu= \dot{\tau}(
-\Delta^{S}_{Q} +(1-\frac{1}{2}P_2(\mu))\Pi^{S}) \, ,
\end{array}\end{equation} 
where $\Pi^{S}=\Delta^{S}_{I2}+\Delta^{S}_{Q0}+\Delta^{S}_{Q2}$.

\subsubsection{Vector Perturbations}

Unlike scalar perturbations, $U$ does not decouple from $I$ and
$Q$ in the vector case, but this can be dealt with very easily.  First we
consider the contribution from the metric:
\begin{equation}
\hat{n}_i \hat{n}_j\dot{h}_{ij}^{vector}=2\mu\sqrt{1-\mu^2}(\dot{h}^V_1
\cos{\varphi} + \dot{h}^V_2 \sin{\varphi}) \, ,
\end{equation}
where 
$\cos{\varphi}=\mathbf{\hat{n}\cdot\mathbf{e_1}}$ and
$\sin{\varphi}=\mathbf{\hat{n}\cdot\mathbf{e_2}}$.
The polarization term is given by \small
\begin{equation}
\mathbf{\Delta}_P^{V\mathcal{S}}(\mu) =
\frac{3}{16\pi}\int \sqrt{1-\mu^2}\sqrt{1-\mu^{\prime2}}
M^V(\mu, \mu^{\prime},\vartheta)
\mathbf{\Delta}_P^{V}(\mu^\prime) d\Omega^{\prime} \, ,
\end{equation} \normalsize
where $\vartheta=\varphi^\prime -\varphi$, and the scattering matrix
$M^V(\mu, \mu^{\prime},\vartheta)$ is given by:
\begin{equation}
\left(\begin{array}{ccc}2\mu\mu^\prime\cos{\vartheta} &
2\mu\mu^\prime\cos{\vartheta} & \mu\sin{\vartheta}\\ 2\mu\mu^\prime\cos{\vartheta} &
2\mu\mu^\prime\cos{\vartheta} & \mu\sin{\vartheta}\\ -2\mu^\prime \sin{\vartheta} &
-2\mu^\prime \sin{\vartheta} & \cos{\vartheta} \end{array}\right) \, .
\end{equation}
Unlike the scalar case, there is a dependence on the
azimuthal angle.  To eliminate it, we introduce new variables defined
as follows:
\begin{equation}\begin{array}{l}
\Delta^{V}_I = -\dot{\imath}\sqrt{1-\mu^2}(\Delta_{I}^{V1}\cos{\varphi}+
\Delta_{I}^{V2}\sin{\varphi}) \\
\Delta_Q^{V} =\mu \sqrt{1-\mu^2}(\Delta_{Q}^{V1}\cos{\varphi}+
\Delta_{Q}^{V2}\sin{\varphi}) \\
\Delta_U^{V} = \sqrt{1-\mu^2}(-\Delta_{U}^{V1}\sin{\varphi}
+\Delta_{U}^{V2}\cos{\varphi}) \, .
\end{array}\end{equation}

With these variables, the equations for each components decouple and
depend
only on $\mu$  so that we can decompose the new variables
into Legendre polynomials.  The equations then read
\begin{equation}\label{trans_v}\begin{array}{l}
\dot{\Delta}^{Vi}_{I} +\dot{\imath}k\mu\Delta^{Vi}_{I} 
+4\dot{\imath}\mu\dot{h}^{V}_i
=-\dot{\tau}(\Delta^{Vi}_{I}-4v_{bi}^{V} 
+\dot{\imath}\mu\Pi^{Vi})\,,\\
\dot{\Delta}^{Vi}_{Q} +\dot{\imath}k\mu\Delta^{Vi}_{Q} 
=-\dot{\tau}(\Delta^{Vi}_{Q} -\Pi^{Vi})\,, \\
\Delta^{Vi}_{Q}+\Delta^{Vi}_{U}=0 \, ,
\end{array}\end{equation}
where
$\Pi^{Vi} = \frac{3}{10}\Delta^{Vi}_{I1} + \frac{3}{10}\Delta^{Vi}_{I3}
+ \frac{13}{20}\Delta^{Vi}_{Q0} -
\frac{3}{4}\Delta^{Vi}_{Q2} + \frac{6}{35}\Delta^{Vi}_{Q4}$.

\subsubsection{Tensor Perturbations}

Again, the Stokes parameter $U$ cannot be set to zero, but, as with
vector perturbations, we will show that can be treated with $Q$.  The
metric term is
\begin{equation}
\hat{n}_i \hat{n}_j\dot{h}_{ij}^{tensor}=(1-\mu^2)(\dot{h}^{T}_{+}
\cos{2\varphi} + \dot{h}^{T}_{\times} \sin{2\varphi}) \, .
\end{equation}

The polarization term is given by
\begin{equation}
\mathbf{\Delta}^{T\mathcal{S}}(\mu) = \frac{3}{32\pi}\int
M^T(\mu,\mu^{\prime},\vartheta)
\mathbf{\Delta}^T(\mu^{\prime}) d\Omega^{\prime} \, ,
\end{equation}
where the scattering matrix $M^T(\mu,\mu^{\prime},\vartheta)$ is given by
\begin{equation}\footnotesize
\left(\begin{array}{ccc}S_-(\mu)S_-(\mu^{\prime})\cos{2\vartheta} & 
-S_-(\mu)S_+(\mu^{\prime})\cos{2\vartheta} &
-\mu^{\prime}S_-(\mu)\sin{2\vartheta} \\
-S_+(\mu)S_-(\mu^{\prime})\cos{2\vartheta} & 
S_+(\mu)S_+(\mu^{\prime})\cos{2\vartheta} &
\mu^{\prime}S_+(\mu)\sin{2\vartheta} \\
2\mu S_-(\mu^{\prime})\sin{2\vartheta} & -2\mu S_+(\mu^{\prime})\sin{2\vartheta}&
2\mu\mu^{\prime}\cos{2\vartheta}\end{array}\right) \, ,
\end{equation}\normalsize
where $S_{\pm}(\mu) = 1\pm\mu^2$.

As with the vector case, we can eliminate the $\varphi$ dependence by
making a change of variables.  In the tensor case, these new variables
are defined as follows:
\begin{equation}\begin{array}{l}
\Delta^{T}_I = (1-\mu^2)(\Delta_{I}^{T+}\cos{2\varphi}+
\Delta_{I}^{T\times}\sin{2\varphi})\,, \\
\Delta_Q^{T} = (1+\mu^2)(\Delta_{Q}^{T+}\cos{2\varphi}+
\Delta_{Q}^{T\times}\sin{2\varphi})\,, \\
\Delta_U^{T} = 2\mu(-\Delta_{U}^{T+}\sin{2\varphi}
+\Delta_{U}^{T\times}\cos{2\varphi}) \, .
\end{array}\end{equation}

Exactly as for vectors, the equations for each polarization
decouple and depend only on $\mu$ so that we can expand them in
Legendre polynomials and carry out the integration.  The resulting
equations are:
\begin{equation}\label{trans_t}\begin{array}{l}
\dot{\Delta}^{T\epsilon}_{I} +\dot{\imath}k\mu\Delta^{T\epsilon}_{I} 
+2\dot{h}^{T}_{\epsilon}
=-\dot{\tau}(\Delta^{T\epsilon}_{I} +\Pi^{T\epsilon})\,,\\
\dot{\Delta}^{T\epsilon}_{Q} +\dot{\imath}k\mu\Delta^{T\epsilon}_{Q} 
=-\dot{\tau}(\Delta^{T\epsilon}_{Q} -\Pi^{T\epsilon})\,,\\
\Delta^{T\epsilon}_{Q}+\Delta^{T\epsilon}_{U}=0 \, ,
\end{array}\end{equation}
where $\Pi^{T\epsilon}=\frac{1}{10}\Delta^{T\epsilon}_{I0}+
\frac{1}{7}\Delta^{T\epsilon}_{I2}+\frac{3}{70}\Delta^{T\epsilon}_{I4}-
\frac{3}{5}\Delta^{T\epsilon}_{Q0}+\frac{6}{7}\Delta^{T\epsilon}_{Q2}-
\frac{3}{70}\Delta^{T\epsilon}_{Q4}$ and $\epsilon=+$ or $\times$.

\subsubsection{Physical significance of moments}\label{moments}

If we use the decomposition (\ref{legendre_exp})
and the recursion relation
\begin{equation}\label{legendre_recur}
\mu P_\ell(\mu)=\frac{1}{2\ell+1}(\ell P_{\ell -1}(\mu)+ (\ell +1)P_{\ell
+1}(\mu)) \, ,
\end{equation}
we obtain the following equations for the moments:
\begin{equation}\label{s_hier}\begin{array}{l}
\dot{\Delta}^S_{I0}=-k\Delta^S_{I1}-\frac{2}{3}\dot{h}\,,\\
\dot{\Delta}^S_{I1}=-\frac{k}{3}(2\Delta^S_{I2}-\Delta^S_{I0})
-\dot{\tau}(\Delta^S_{I1}-\frac{4}{3k}\theta_{b})\,,\\
\dot{\Delta}^S_{I2}=\frac{k}{5}(2\Delta^S_{I1}-3\Delta^S_{I3})
-\dot{\tau}(\Delta^S_{I2} - \frac{1}{10}\Pi^S)-\frac{4}{15}\dot{h}^S\,,\\
\dot{\Delta}^S_{I\ell}=\frac{-k}{2\ell +1}(\ell\Delta^S_{I\ell -1}-(\ell
+1)\Delta^S_{I\ell +1})-\dot{\tau}\Delta^S_{I\ell} \; , ~~\ell>2\,,\\
\dot{\Delta}^S_{Q\ell}=\frac{k}{2\ell +1}(\ell
\Delta^S_{Q\ell-1}-(\ell+1)\Delta^S_{Q\ell +1}) 
+ \dot{\tau}(-\Delta^S_{Q\ell} +\Pi^S(\delta_{\ell
0}+\frac{1}{10}\delta_{\ell 2})) \,,
\end{array}\end{equation}
\begin{equation}\label{v_hier}\begin{array}{l}
\dot{\Delta}^{Vi}_{I0}= -k\Delta^{Vi}_{I1}
-\dot{\tau}(\Delta^{Vi}_{I0} +4v_{bi}^V)\,,\\
\dot{\Delta}^{Vi}_{I1}=
\frac{k}{3}(\Delta^{Vi}_{I0}-2\Delta^{Vi}_{I2})
-\frac{4}{3}\dot{h}^V_i -\dot{\tau}(\Delta^{Vi}_{I1} 
+\Pi^{Vi})\,,\\
\dot{\Delta}^{Vi}_{I\ell}= \frac{k}{2\ell
+1}(\ell\Delta^{Vi}_{I\ell -1} -(\ell+1)\Delta^{Vi}_{I\ell +1})
-\dot{\tau}\Delta^{Vi}_{I\ell} \; , ~~\ell>1\,,\\
\dot{\Delta}^{Vi}_{Q\ell}= \frac{k}{2\ell
+1}(\ell\Delta^{Vi}_{Q\ell -1} -(\ell+1)\Delta^{Vi}_{Q\ell +1})
-\dot{\tau}\Delta^{Vi}_{Q\ell}
+\dot{\tau}\Pi^{Vi}\delta_{0\ell} \,,
\end{array}\end{equation}
\begin{equation}\label{t_hier}\begin{array}{l}
\dot{\Delta}^{T\epsilon}_{I0}=
-k\Delta^{T\epsilon}_{I1}-2\dot{h}^T_\epsilon
-\dot{\tau}(\Delta^{T\epsilon}_{I0}+\Pi^{T\epsilon})\,, \\
\dot{\Delta}^{T\epsilon}_{I\ell}=\frac{k}{2\ell +1}
(\ell\Delta^{T\epsilon}_{I\ell -1} -(\ell+1)\Delta^{T\epsilon}_{I\ell
+1}) -\dot{\tau}\Delta^{T\epsilon}_{I\ell} \; , ~~\ell>0\,,\\
\dot{\Delta}^{T\epsilon}_{Q\ell}=\frac{k}{2\ell +1}
(\ell\Delta^{T\epsilon}_{Q\ell -1} -(\ell+1)\Delta^{T\epsilon}_{Q\ell
+1}) -\dot{\tau}\Delta^{T\epsilon}_{Q\ell}
-\dot{\tau}\Pi^{T\epsilon}\delta_{\ell 0} \,.
\end{array}\end{equation}

These moments are related to the energy momentum tensor of
photons and (massless) neutrinos.  First, let us consider the energy
density.  We have that 
\begin{equation}
\delta T_{00}=-\delta\rho= -\frac{\rho}{4\pi}\int\Delta_I d\Omega \, ,
\end{equation} so that 
\begin{equation}
\delta =\Delta_{I0}^S \, .
\end{equation}
Secondly, the velocity perturbations
\begin{equation}
\delta T_{0i}=(\rho +p)v_i=\frac{3p}{4\pi}\int\Delta_I \hat{n}_i
d\Omega \, ,\end{equation}
which yields
\begin{equation}\begin{array}{l}
\theta =\frac{3}{4}k\Delta^S_{I1} \,,\\
v^V_i=\frac{1}{4}(\Delta^{Vi}_{I0}+\Delta^{Vi}_{I2}) \, .
\end{array}\end{equation}
Finally we have the anisotropic stresses
\begin{equation}
\delta T_{ij}-\frac{1}{3}T^k_k\delta_{ij}=p\Sigma_{ij}=\frac{3p}{4\pi}
\int\Delta_I (\hat{n}_i \hat{n}_i -\frac{1}{3}\delta_{ij})d\Omega \, ,
\end{equation}
which yields
\begin{equation}\begin{array}{l}
\Sigma^S = -3\Delta^S_{I2}\,, \\
\Sigma^V_i= \frac{3}{5}(\Delta^{Vi}_{I1}+\Delta^{Vi}_{I3})\,,
\\ \Sigma^T_{\epsilon}=\frac{2}{5}\Delta^{T\epsilon}_{I0}+
\frac{4}{7}\Delta^{T\epsilon}_{I2}+\frac{6}{35}\Delta^{T\epsilon}_{I4}
\, .
\end{array}\end{equation}

We now have an open hierarchy of equations for all the components of
the energy-momentum tensor.  Numerically, it has been found that
truncating the expansion by simply setting the $\ell^{th} $ moment to
$0$ propagates a sizeable error back to the first moment.  It has also
been found~\cite{ma95} that the following expression gives excellent results:
\begin{equation}
\dot{\Delta}_{\ell_{max}}=
k\Delta_{\ell_{max}-1}-
\left(\frac{\ell_{max}-1}{\eta}+\dot{\tau}\right)
\Delta_{\ell_{max}} \, , 
\end{equation}
where $\ell_{max}$ is the moment at which the series is truncated.
The above expression is good for all photon and neutrino equation
hierarchies.

\subsection{Non-Relativistic Matter}

For non-relativistic matter, it is not necessary to consider a full
phase space expansion, as all but the first few moments are totally
negligible.  Our starting point will be the equations derived from
the conservation of the energy-momentum tensor (\ref{t_cons}).

CDM does not couple with other types of matter (except
gravitationally), so we can immediately use the conservation
equations.  Here $w=c^2_s=0$.  Also, in the (comoving) synchronous gauge, CDM has
$\mathbf{v}_c=\Sigma_{ij}=0$, so (\ref{t_cons}) simplifies
to
\begin{equation}
\dot{\delta}_c=-\frac{\dot{h}}{2} \, .
\end{equation}

For baryons, these conservation equations (\ref{t_cons}) become
\begin{equation}\begin{array}{c}
\dot{\delta}_b=-(\theta_b +\frac{\dot{h}}{2})\\
\dot{\theta}_b=-\frac{\dot{a}}{a}\theta_b +c^2_sk^2\delta_b\\
\dot{v}^V_{bi}=-\frac{\dot{a}}{a}v^V_{bi} \, ,
\end{array}\end{equation}
where we have used the fact that $w$,~$c^2_s \ll 1$.
However, baryons interact with photons through Thompson scattering.  So
we must correct the last two equations for momentum exchange between
the two fluids.  The
equations for photons obtained in section~\ref{moments} read
\begin{equation}\begin{array}{c}
\dot{\theta}_{\gamma}=k^2(\frac{1}{4}\delta_{\gamma}
-\frac{1}{6}\Sigma^S)
+\dot{\tau}(\theta_b-\theta_{\gamma})\\
\dot{v}^V_{\gamma i}=\frac{k}{4}\Sigma^V_{\gamma i}
+\dot{\tau}(v^V_{b i}-v^V_{\gamma i}) \, .
\end{array}\end{equation}

By comparing with the conservation equations with
$w=c^2_s=\frac{1}{3}$, we see that there is an extra drag term
$\dot{\tau}(v_{b}-v_{\gamma})$.  Hence, for momentum to be conserved, 
we have to add the term
$\frac{4\rho_{\gamma}}{3\rho_b}\dot{\tau}(v_{\gamma}-v_{b})$ to
the previous two equations, which then become
\begin{equation}\begin{array}{c}
\dot{\theta}_b=-\frac{\dot{a}}{a}\theta_b +c^2_sk^2\delta_b
+R\dot{\tau}(\theta_{\gamma}-\theta_b) \\
\dot{v}^V_{bi}=-\frac{\dot{a}}{a}v^V_{bi}+ R\dot{\tau}(v^V_{\gamma
i}-v^V_{bi}) \, .
\end{array}\end{equation}
where we have defined $R=\frac{4\rho_{\gamma}}{3\rho_b}$.

\subsection{The Sachs-Wolfe Formula}

The temperature anisotropy formula is derived by considering
perturbations in the photon energy $E_{\gamma}$ along the unperturbed
path $X^{\mu}=\hat{n}^{\mu}\eta$.  Here $\mathbf{\hat{n}}$ is the line of sight
direction (i.e. pointing in the direction opposite to the photon's
travel).  For a proof, see e.g.~\cite{sachs67,pen94}.
In the synchronous gauge, the temperature fluctuations are given by:
\begin{equation}
\frac{\delta T}{T} = \frac{\delta_{\gamma}}{4} -
\mathbf{v_{\gamma}\cdot\hat{n}}
-\frac{1}{2}\int\dot{h}_{ij}\hat{n}_i\hat{n}_j d\eta \, .
\end{equation}

The above formula was derived under the assumption of instantaneous
recombination.  To treat the finiteness of the surface of
last-scattering, we integrate the expression for the temperature
fluctuations over the probability of free-streeming for a photon,
$e^{-\tau}d\tau$:
\begin{equation}\label{sw_corr}
\overline{\frac{\delta T}{T}} = \int_0^{\eta_0}\frac{\delta
T}{T}\dot{\tau}e^{-\tau}d\eta = 
\int_0^{\eta_0} \left(\dot{\tau}e^{-\tau}\left(\frac{\delta_{\gamma}}{4} -
\mathbf{v_\gamma}\cdot\hat{n}\right)
-\frac{1}{2}e^{-\tau}\dot{h}_{ij}\hat{n}_i\hat{n}_j \right)d\eta \, ,
\end{equation}
where the ISW term was obtained by integrating by parts and setting
the surface term to zero because the visibility function is very
sharply defined around the time of recombination and hence is utterly
negligible at $\eta=0$ or today.  

The calculation of the thermal history of the universe for an 
arbitrary set of cosmological parameters was achieved with an 
integrated package which will be described in more detail 
elsewhere \cite{thesis}.  The Friedmann equations were solved
simultaneously with the ionization rate equations for hydrogen and 
helium.  The results were compared for accuracy against RECFAST 
\cite{seager99} for which the code provided an independent 
check
\cite{thesis}.  These computations start very deep in the
radiation era and end today in order to create high accuracy tables
from which the relevant quatities for the Boltzmann evolution, such as
the opacity and visibilty function, are later interpolated using a
cubic spline.

\section{Cosmic defect source terms}\label{sec-sources}

The perturbation source terms in the Boltzmann evolution are given 
by the Fourier transform of the cosmic defect energy-momentum 
tensor, decomposed into scalar, vector and tensor components. 
The code can accept the energy-momentum tensor of any set of 
`active' sources with appropriate initial conditions, 
whether these are cosmic strings, global defects or other 
more exotic phenomena.  We have experimented with inputting 
from global defect simulations, but the focus of our attention 
here is on local cosmic string simulations which inherently can
achieve far higher resolution and greater dynamic range. 

The cosmic string simulations were performed using the AS
code~\cite{allen90}, for which the methods employed and key results
have been described in detail elsewhere.
The strings are approximated 
by a two-dimensional worldsheet defined by $x^\mu_s(\sigma,\,t) = 
(t,{\bf x}_s(\sigma,\, t))$, 
where the position
${\bf x}_s$ is a function of the two coordinates 
$\sigma$ (spacelike) and conformal time $t$ in an unperturbed
FRW background.  We can impose 
the condition that the velocity $\dot{\bf x}$ is transverse to 
the tangent vector 
${\bf x}'_s \equiv d{\bf x}_s/d\sigma$ along the string, that is, 
$\dot {\bf x}_s\cdot {\bf x}_s'=0$.  The strings are evolved by 
splitting the equations of motion for the strings into their 
characteristic left-
and right-moving modes, which are damped slightly by the expansion of 
the universe.  The key points to note about the string network simulations are
that above a minimum resolution, energy conservation is accurately 
satisfied during the numerical evolution to within a fraction of 
one percent, and over a dynamic range approaching an order of magnitude
in conformal time (that is, several decades in redshift).  

The energy-momentum 
tensor of the strings is given by 
\begin{equation}\label{stringem}
\Theta^{\mu\nu}\sqrt{-g} = \mu_s\int
d\sigma\,(\epsilon\dot x_s^{\mu}\dot x_s^{\nu} - 
\epsilon^{-1}x_s^{\prime
\mu}x_s^{\prime \nu})\,\delta^{(3)}(\mathbf{x}^{\sigma} -
\mathbf{x}_s^{\sigma}(\sigma)) \, ,
\end{equation}
where $\mu_s$ is the linear energy density of the string, 
$\epsilon = \left({\bf x}'^2/(1-\dot{\bf x}^2)\right)^{1/2}$, 
and $g$  is the determinant of the background FRW metric ($\sqrt{-g}=  a^4$).
All the components of $\Theta_{\mu\nu}(\mathbf{x},\eta)$
 were calculated at each 
point on the string network and then interpolated onto a high resolution 
grid.  Since there is an implicit differentiation of the spatial 
distribution of (\ref{stringem}) in the scalar-vector-tensor 
decomposition described in section~\ref{sec_SVT}, 
the interpolation must be sufficiently smooth for the decomposition 
to remain well behaved.  Simple cloud-in-cell interpolation, for example, 
produced poorly controlled results, so we chose a two step approach with 
a higher order scheme.  The first step was to use the 
triangular shaped cloud (TSC) interpolation involving the 
27 nearest neighbour points of the string segment. The weight function 
is given by 
\begin{eqnarray}
W_3(x) = \cases{ \frac{3}{4} - x^2 \,, &$|x|<\frac{1}{2}\,,$\cr \nonumber
\frac{1}{2}\left(\frac{3}{2} - x^2\right)^2 \,, 
& ${\frac{1}{2}}<|x|\leq\frac{3}{2}\,,$\cr}
\end{eqnarray}
where $|x|$ is the distance between grid point and string in the
$x$-direction.  This weight function is multiplied by the appropriate weights 
in $y$- and $z$-directions.  The
second step is to smooth the result further by employing a gaussian 
window function in Fourier space, that is, proportional to $\exp
(-k^2/k_R^2)$ with $k_R^{-1}$ corresponding to about three grid points;
the discrete Fourier transform employed the NAG routine \texttt{C06PXF}.
Energy-momentum conservation was closely monitored and maintained to 
within a fraction of one percent by this process. 

The smoothing from the interpolation and filtering leads to an
apparent loss of resolution, but this need not be important.  A
well-behaved decomposition can be obtained at very high resolution on
a larger grid and then used at lower resolution for the Green's
function integration.  We note also that, in Fourier space, the
decomposed scalar, vector and tensor parts of (\ref{stringem})
calculated as in section~\ref{sec_SVT} can be stored on only half the
complex grid, because the Fourier transform of a real quantity
satisfies $f(\mathbf{k}) = f^*(-\mathbf{k})$, it is necessary to treat
only just over half of the grid, that is, $i_x$, $i_y = 0, 1, ...,
N-1$ and $i_z=0, 1, ..., N/2$ (it is in fact possible to use exactly
half, but the relationship between components is more
complicated~\cite{NRiC2}).

\begin{figure}[t]
\resizebox{8cm}{!}{\includegraphics{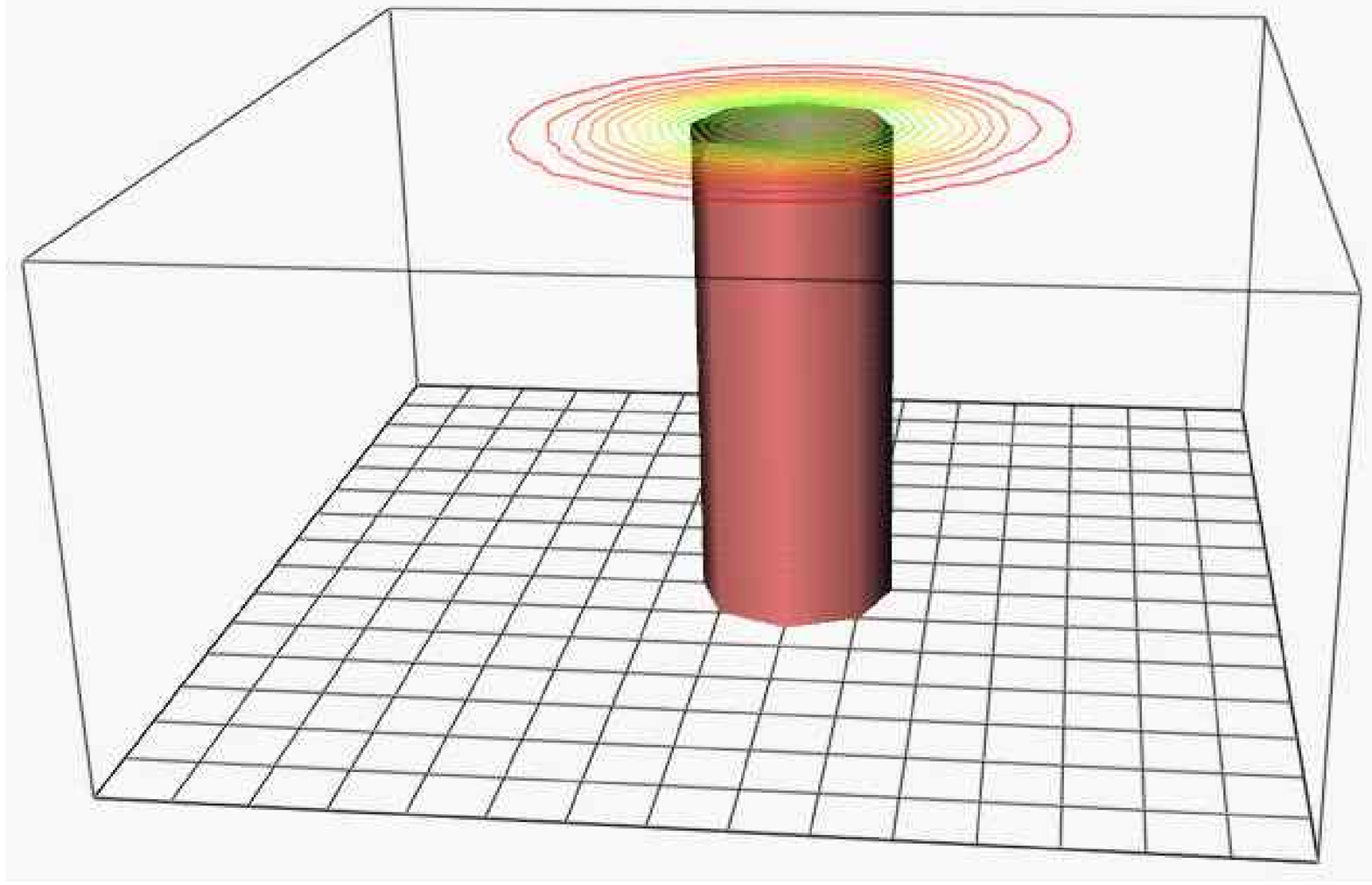}}
\resizebox{8cm}{!}{\includegraphics{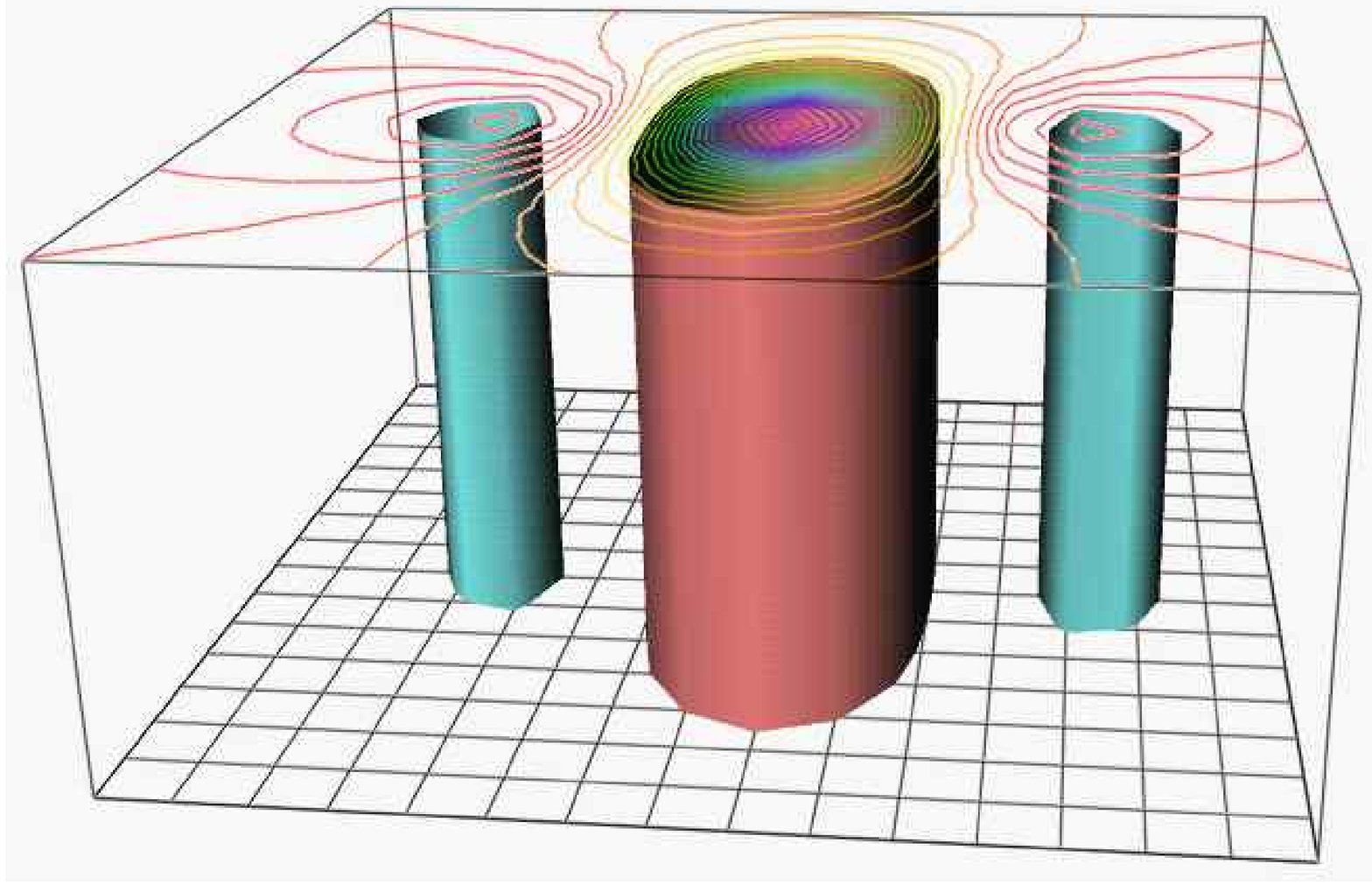}}
\resizebox{8cm}{!}{\includegraphics{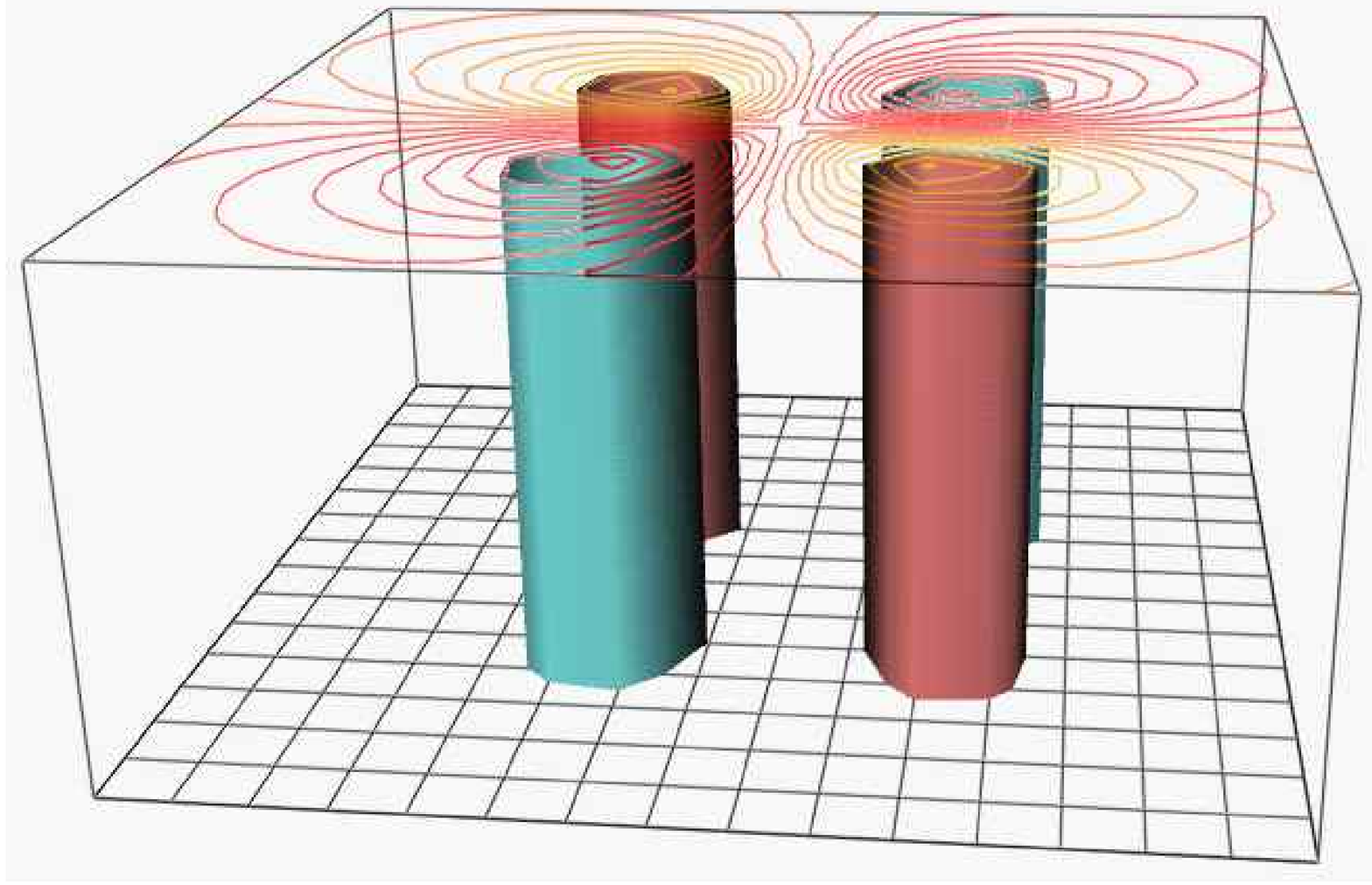}}
\resizebox{8cm}{!}{\includegraphics{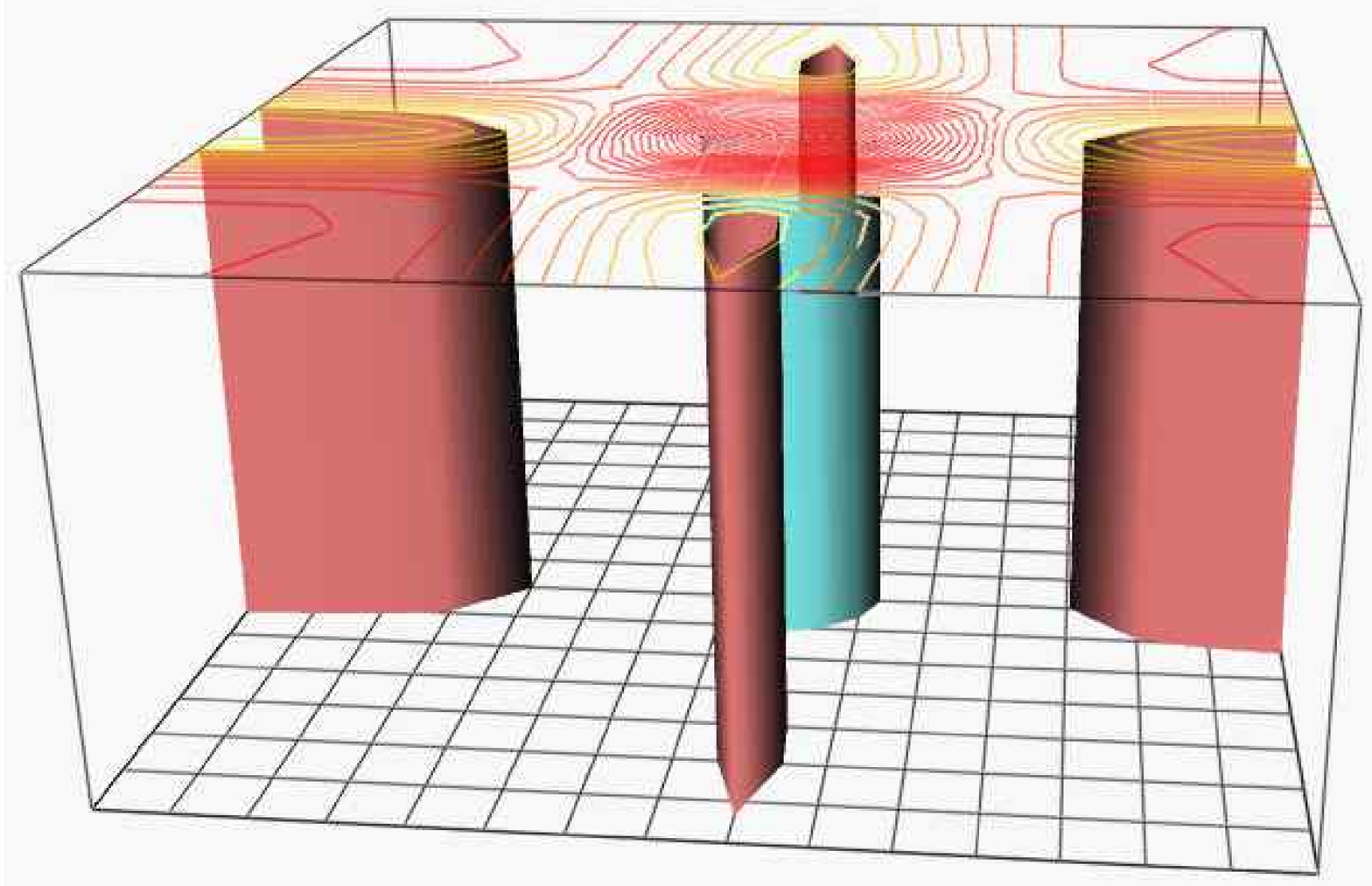}}
\caption{Positive and negative isosurfaces of scalar, vector and
tensor components in real space created from the decomposition of the
energy-momentum tensor of a straight string.  Clockwise from top-left
$\Theta_{00}$ (positive), $\Theta_S$ (positive centre and negative
sidelobes), $\Theta_T^{+}$ (negative centre and positive sidelobes)
and $\Theta_V^{1}$ (alternating negative and positive). Also shown are
contours in a plane transverse to the string.}\label{fig-svt}
\end{figure}

Figure~\ref{fig-svt} illustrates the decomposed scalar, vector and
tensor parts of (\ref{stringem}) for a straight cosmic string
calculated as in section~\ref{sec_SVT}.  The scalar $\Theta_{00}$ is
localised in real space, but the scalar and vector projection
operators acting on the tensor components $\Theta_{ij}$ yields the
apparently non-local results shown.  This demonstrates that the
decomposed components must be calculated and evolved with great care;
the final temperature pattern must reflect the locality (or causality)
of the sources which generated them.

\section{Numerical Implementation}\label{sec-numerics}

The various evolution equations derived in section~\ref{sec_pert} can be
written as three systems (one for scalar, vector and tensor) of the
following form:
\begin{equation}\label{sys_ode}
\frac{d\mathbf{y}}{d\eta} = \mathcal{A}(k,\eta)\mathbf{y} +
\mathbf{q}(\mathbf{k},\eta) \, ,
\end{equation}
where $\mathbf{y}$ and $\mathbf{q}$ are vectors of dimension
$n_{var}$, the number of equations to solve and $\mathcal{A}$ is a real square
matrix of the same order. 
The components of $\mathbf{y}$ are the metric
perturbation, CDM and baryon density contrasts and peculiar
velocities and photon and neutrino moments and those of $\mathbf{q}$
are the defect source terms. Given that there is a continuous
contribution at all times from these sources throughout space, we
cannot simply solve this problem by projecting our initial conditions
forward to the present time with transfer functions (as in inflation).
However, rather than directly solving the differential equations at 
every grid point, it is much more efficient to employ a Green's function 
approach.

In order to take advantage of the fact that the evolution equations
depend only on the magnitude of the wavevector and not its
orientation, we proceed to solve (\ref{sys_ode}) by constructing the
fundamental matrix $Y$ of the system (e.g,
see~\cite{aschor98}) which satisfies
\begin{equation}\label{sys_homo}
\begin{array}{c}
\displaystyle\frac{d}{d\eta}Y = \mathcal{A}Y \,,\\
Y(0) = \openone \, ,
\end{array} \end{equation}
where $\openone$ is the identity matrix.  Now, given some initial conditions
\begin{equation}
\mathbf{y}(0) = \mathbf{c}\, ,
\end{equation}
the solution to (\ref{sys_ode}) is
\begin{equation}\label{solution}
\mathbf{y}(\eta) = Y(\eta)\left(\mathbf{c} + \int_0^{\eta}
Y^{-1}(\eta^{\prime}) \mathbf{q}(\eta^{\prime}) d\eta^{\prime}
\right) \, .
\end{equation}
The above equation reveals a similarity between the matrix $Y^{-1}$
and a Green's function.  However, the latter nomenclature is not 
strictly 
appropriate in this case, as one of the basic properties of a Green's
functions is undefined in a first order problem.

\subsection{Constructing the Matrices}

Since the system (\ref{sys_ode}) is first order, we need to reexpress
the metric equations in first order form (all the matter equations are
already first order).
For the scalar metric equations, we use the two equations of motion,
with the first constraint equation to replace the $k^2h^-$ term, to
obtain the following first order equations for $\dot{h}$ and
$\dot{h}^S$:
\begin{equation}\begin{array}{c}
\displaystyle\frac{d\dot{h}}{d\eta} = -\frac{\dot{a}}{a}\dot{h} -8\pi
Ga^2(\delta\rho +3\delta p) -8\pi G(\Theta_{00} + \Theta) \,,\\
\displaystyle\frac{d\dot{h^S}}{d\eta} = \frac{\dot{a}}{a}(\dot{h}-2\dot{h^S}) +8\pi
Ga^2(2p\Sigma^S - \delta\rho) + 8\pi G(2\Theta^S - \Theta_{00})\,.
\end{array}\end{equation}
The vector metric equation of motion is already in a form that is 
first order for $\dot{h}^V$:
\begin{equation}
\frac{d\dot{h^V}}{d\eta} = -2\frac{\dot{a}}{a}\dot{h}^V
+ 16\pi Ga^2p\Sigma^V + 16\pi G\Theta^V \, .
\end{equation}
For the tensor modes, we use the standard  
order reduction, by considering $h^T$ and $\dot{h}^T$ as
separate variables:
\begin{equation}
\begin{array}{c}
\displaystyle\frac{dh^T}{d\eta} = \dot{h}^T \,,\\
\displaystyle\frac{d\dot{h}^T}{d\eta} = -2\frac{\dot{a}}{a}\dot{h}^T
-k^2h^T + 16\pi Ga^2p\Sigma^T + 16\pi G\Theta^T \, .
\end{array}\end{equation}

To construct $Y$, we numerically solve (\ref{sys_homo}), which
involves solving 3 times (scalar - vector - tensor) $n_{var}$ systems
of $n_{var}$ coupled equations for a chosen number  $M$ of 
the wavenumbers $k$ (typically $M\sim N$ for an $N^3$ grid).  This contrasts
with solving 10 systems (real and imaginary parts of 1 scalar, 2
vectors and 2 tensors) of $n_{var}$ equations for $N^3/2$ wavevectors
$\mathbf{k}$ if we were to solve (\ref{sys_ode}) directly at every
grid point.  Hence this Green's function approach
represents a reduction of computing time by a factor
of $5N^2/3n_{var}$.  Since $n_{var}\simeq 40$, this factor is more
than 1,000 for a 256$^3$ box and 10,000 for a 1024$^3$ box.  Of course,
there are extra computations involved: the inversion of the matrices
(for which we use the NAG \texttt{F07ADF} and \texttt{F07ADJ} Fortran
routines) and the actual integration of (\ref{solution}), but this is
insignificant compared with the time taken to solve the extra
equations.  In fact, this method is so efficient it is constrained by 
the amount of memory (or disk space) required.

\begin{figure}[t]
\noindent
\hspace{1.5cm}
\resizebox{12cm}{!}{\includegraphics{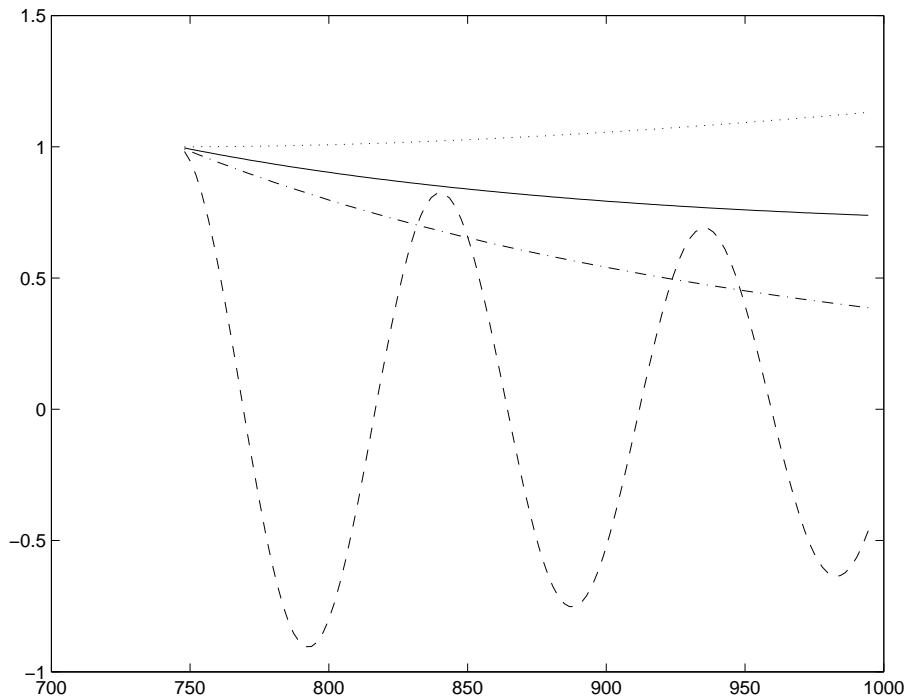}}
\caption{Time evolution of elements of the fundamental
matrices for $\eta = 3\eta_{dec}$ to $4\eta_{dec}$.  These particular
diagonal elements are those associated with ISW effects, i.e. the time
derivative of the metric terms: scalar trace $\dot{h}$ (solid),
anisotropic scalar $\dot{h}^S$ (dot-dashed), which coincides with the
vector $\dot{h}^V$ and tensor $\dot{h^T}$ (dashed).  Also illustrated
is the CDM density contrast $\delta_c$ (dotted).  The conformal time
along the horizontal axis is given in Mpc}\label{yeta_fig}
\end{figure}

\begin{figure}[t]
\resizebox{12cm}{!}{\includegraphics{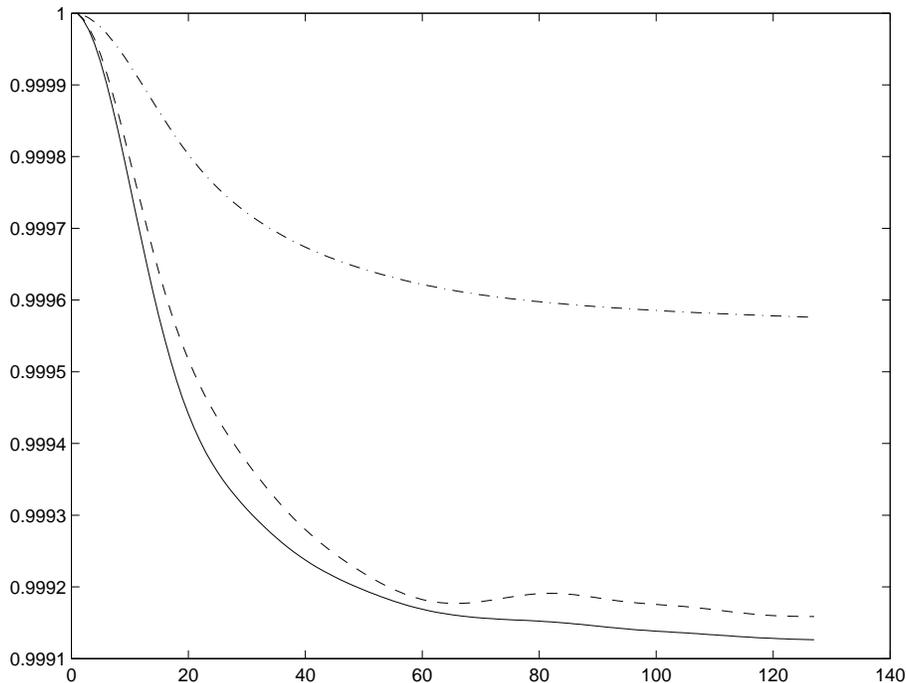}}
\caption{Diagonal scalar matrix elements $Y_{11}$, $Y^{-1}_{11}$ and $Y_{33}$ from 
the late matter era corresponding respectively to $\dot h$ (continuous
and dashed lines) and $\delta_c$ (dot-dashed).  The normalisation is
arbitrary and the $k$ scale is $1.8\times10^{-4}Mpc^{-1}$.}\label{Ys}
\end{figure}

To numerically solve the perturbation equations, we use
\texttt{dverk.f}, a sixth order Runge-Kutta Fortran routine.
Figure~\ref{yeta_fig} shows the time evolution for some diagonal
elements of the fundamental matrices.
Our Boltzmann code was extensively tested for inflationary scenarios
and found to be in excellent agreement with publicly available
codes~\cite{thesis}.
The matrix elements are computed for $k=0$ to $k=1.8N\Delta k/2$ with a
spacing of $0.9\Delta k$, where
$\Delta k$ is the simulation grid spacing.  The value of $1.8$ was
chosen to be sightly higher than the maximum possible value
$\sqrt{3}$.  This is to ensure that there is no need to extrapolate
the matrix elements.  This $k$-spacing is sufficiently dense for the
elements of the matrices and their inverses to be linearly interpolated
for the appropriate value of $k$ at every simulation grid point
(see FIG.~\ref{Ys}).

Since only a few components of the vector $\mathbf{q}$ are
non-zero (two for scalars, one for vectors and one for tensors), only
the corresponding rows of $Y^{-1}$ need to be stored.  Also, since
only 10 quantities are needed to compute the SW integral
($\delta_{\gamma}$, $\theta_{\g}$, $v_{\g}^V$, $h$, $h^S$, $h^V$ and $h^T$),
only the corresponding columns of $Y$ need to be stored in principle.
In practise however, all components $y$ need to be kept to allow
the computations to be made in several stages.  There are two reasons
why we want to do this.  The main reason is that, even though all the
components of the fundamental matrices are well behaved at all times,
the evolution results in different components having very large
ratios.  This results in $Y$ become non-invertible \textit{to machine
precision}.  This happens more often at early times and for scalar
modes.  The solution (\ref{solution}) is then computed by
\begin{equation}\begin{array}{c}
\mathbf{y}(\eta) = Y^{(i)}(\eta)\left(\mathbf{y}(\eta_i) +
\int_{\eta_i}^{\eta_{i+1}}
Y^{(i)~-1}(\eta^{\prime}) \mathbf{q}(\eta^{\prime}) d\eta^{\prime}
\right) \\
\eta_i < \eta < \eta_{i+1} \, , ~~~i = 0, 1, ... n_{stage} \\
\mathbf{y}(\eta_{i=0}) = \mathbf{c} \, ,
\end{array}\end{equation}
where the corresponding matrices $Y^{(i)}$ and $Y^{(i)~-1}$ are
computed from $\eta_i$ to $\eta_{i+1}$.  The second reason for wanting
to allow the possibility of running the computations in several stages
is more practical: this enables us to checkpoint the code in order
minimize the effects of a system crash.

\subsection{Initial Conditions}\label{compensation}

As our simulations start at a time much later than that of the phase
transition that created the defects, we must specify initial
conditions that are consistent with a fully formed network.  In any
case, the local string simulations used are based on an effective action and
hence cannot be used to simulate the phase transition itself.

One of the nice things about the solution (\ref{solution}) is that it
is expressed as a sum of a term depending on the initial conditions
and a second one depending on the defect sources.  This provides us
with an easy way to assess the relative importance of the initial
conditions and hence to see if our results are sensitive or not to
them.

One possible choice is to set all gradient terms and time derivatives
to zero in the constraint equations.  This leads to the following
initial conditions:
\begin{equation}\begin{array}{c}
\delta_c = \delta_b = \frac{3}{4}\delta_{\g} = \frac{3}{4}\delta_{\nu} =
-\frac{\Theta_{00}}{a^2(\rho + p)}\,, \\
\theta_b = \theta_{\g} = \theta_{\nu} = \frac{\Theta_D}{a^2(\rho +
p)}\,,
\\
v_b^V = v_\g^V = v_\nu^V = \frac{{\mathcal P}^V}{a^2(\rho + p)} \, ,
\end{array}\end{equation}
the last line being valid for all three components.  This choice
is consistent with setting the pseudo-energy $\tau_{00}$ and pseudo-momenta
$\tau_{0i}$ to zero, a set of initial conditions consistent with 
matching an instantaneous defect-forming phase transition to a 
homogeneous initial state (see, for example,
\cite{veeraraghavan90,amery2002}).

\subsection{Integration over the String Energy-Momentum Tensor}

The matrix elements are integrated
over the energy-momentum tensor using the trapezoidal rule.
The equations being real and scalar, the same matrices are used for both the
real and imaginary parts as well as the two vector and two tensor
components, so that the integration is repeated twice for scalars, and
four times for both vectors and tensors.

The grids of the quantities needed to compute the SW integral,
$\delta_{\gamma}$, $\mathbf{v}_{\gamma}$ and $h_{ij}$ are then
completed at each timestep by complex conjugation and Fourier
transformed back to real space.  They are then projected onto the end
of the vectors $\hat{n}_i(\eta - \eta_0)$, where the $\hat{n}_i$ are
the pixel directions, using the inverse cloud-in-cell scheme:
\begin{equation}\begin{array}{l}
f(x,y,z) =  w_x(w_y(w_zf(x_1,y_1,z_1)
+(1-w_z)f(x_1,y_1,z_2))  \\
\mbox{} +(1-w_y)(w_zf(x_1,y_2,z_1) 
+(1-w_z)f(x_1,y_2,z_2)))  \\
\mbox{} +(1-w_x)(w_y(w_zf(x_2,y_1,z_1)
+(1-w_z)f(x_2,y_1,z_2))  \\
\mbox{} +(1-w_y)(w_zf(x_2,y_2,z_1)
+(1-w_z)f(x_2,y_2,z_2))) \, ,
\end{array}\end{equation}
where the $x_1$, $x_2$, ... define the grid element and the weights
are given by $w_x = 1 - x_1$ and similarly for the other two.  The
temperature fluctuations are then computed using (\ref{sw_corr}).  This
integration is also performed using the trapezoidal rule as the
integrand varies very smoothly along each line of sight.

\section{Test of the Numerics}

In order to test the validity of our methods, we compare the result of
simulations with analytic results.  One of the difficult aspects of
such comparisons is that analytic results are mostly obtained for
defect configurations (i) in Minkowski space and (ii) in the no matter
approximation.  In an expanding Universe, with a complete treatment of
matter perturbations, these approximations would be relatively good
only in the late matter era and on scales much smaller than the
horizon.

\subsection{Kaiser-Stebbins Effect}

To illustrate this, we consider an infinite straight string moving in
a direction perpendicular to the line-of-sight.  According
to~\cite{kaiser84b}, this will produce a discontinuity in the
temperature fluctuations.  This calculation is done in the limit that
a plane wave of CMB radiation is propagating in the direction of the
observer.  Objects behind the string receive a boost towards the plane
in which the string is moving because of the gravitational effect of
its deficit angle $\Delta = 8\pi G\mu$.  This means that the CMB
photons that were behind the string when they cross its plane will be
blueshifted, so that they will be hotter than those that were in front
of the string.  The magnitude of the discontinuity is~\cite{gott85}:
\begin{equation}\label{kseffect}
\frac{\Delta T}{T} = 8\pi G\mu v\gamma \, ,
\end{equation}
where $v$ is the string velocity and $\gamma$ is the Lorentz factor.

This provides an excellent test of our map-making pipeline,
particularly the vector modes, which cannot be compared with
inflationnary results such as is done for the scalars and tensors in
chapter~\cite{thesis}.  However, in a realistic
expanding Universe, such as those studied in this dissertation, this
effect is more complicated due to the presence of matter, the growth (or
decay) of SVT components, the curvature of the microwave sky and the
decelaration of the string.  In addition, this idealised solution must
be studied in a periodic box with causal effects due to the limited
dynamic range.

\begin{figure}
\resizebox{7.5cm}{7.5cm}{\includegraphics{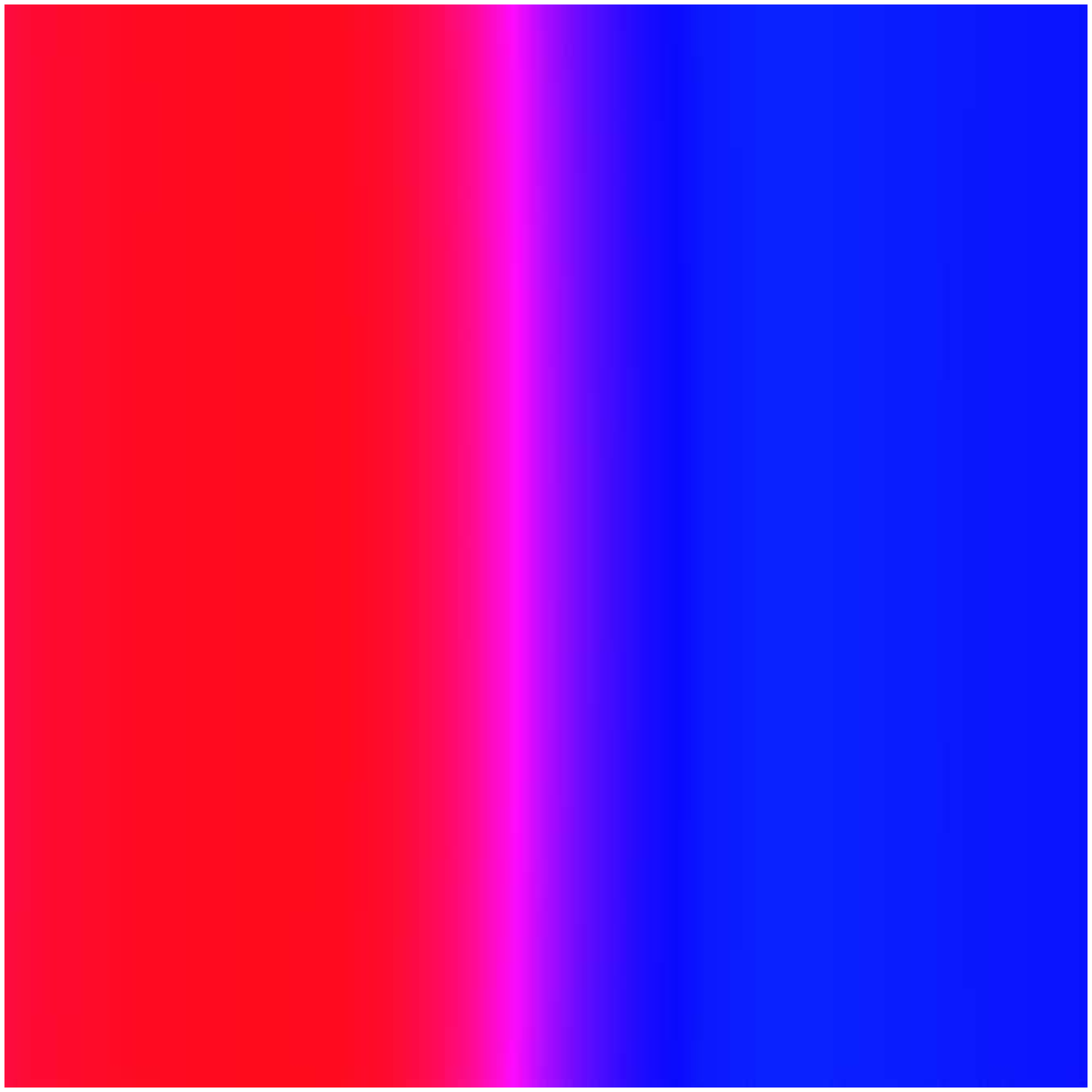}}
\resizebox{7.5cm}{7.5cm}{\includegraphics{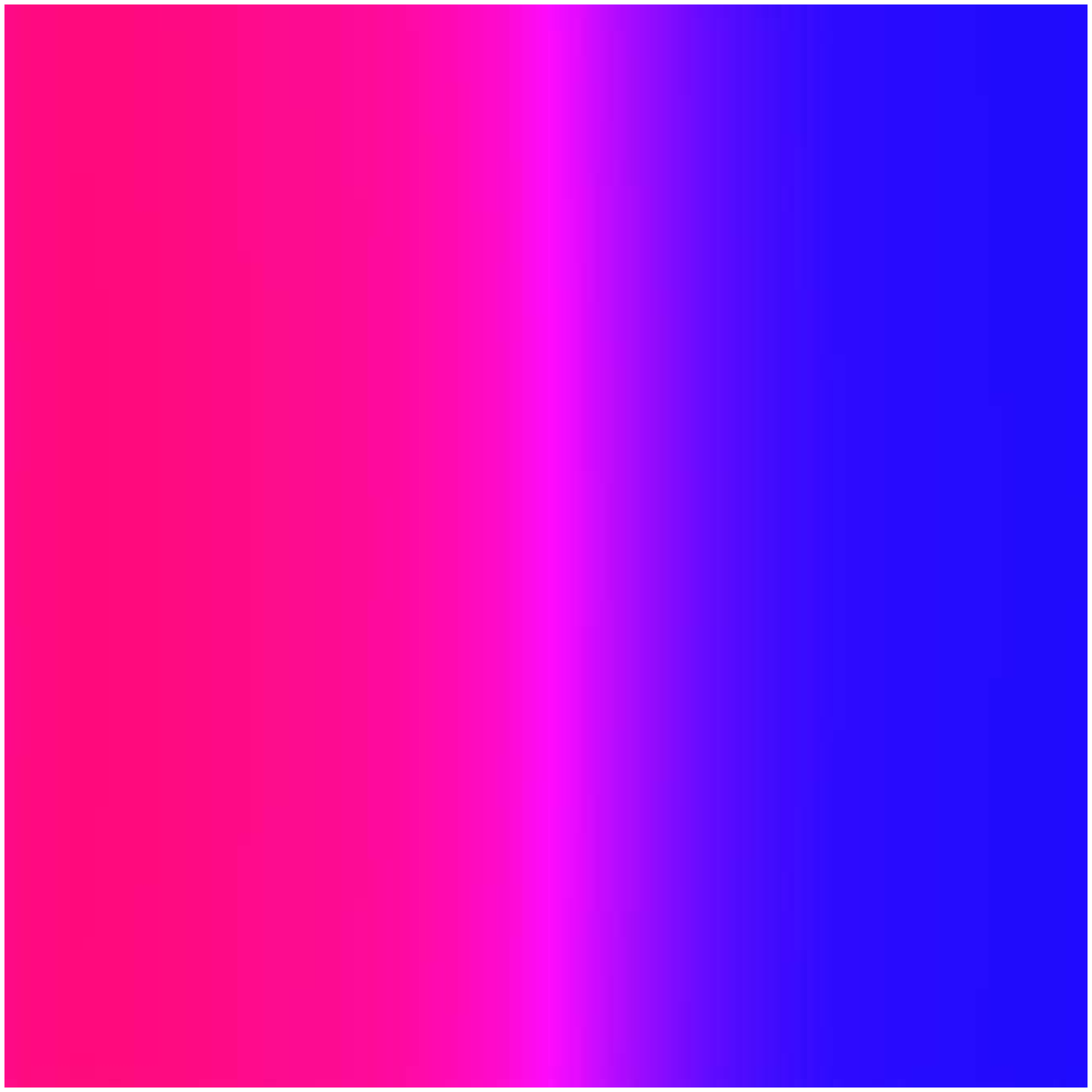}}
\resizebox{7.5cm}{7.5cm}{\includegraphics{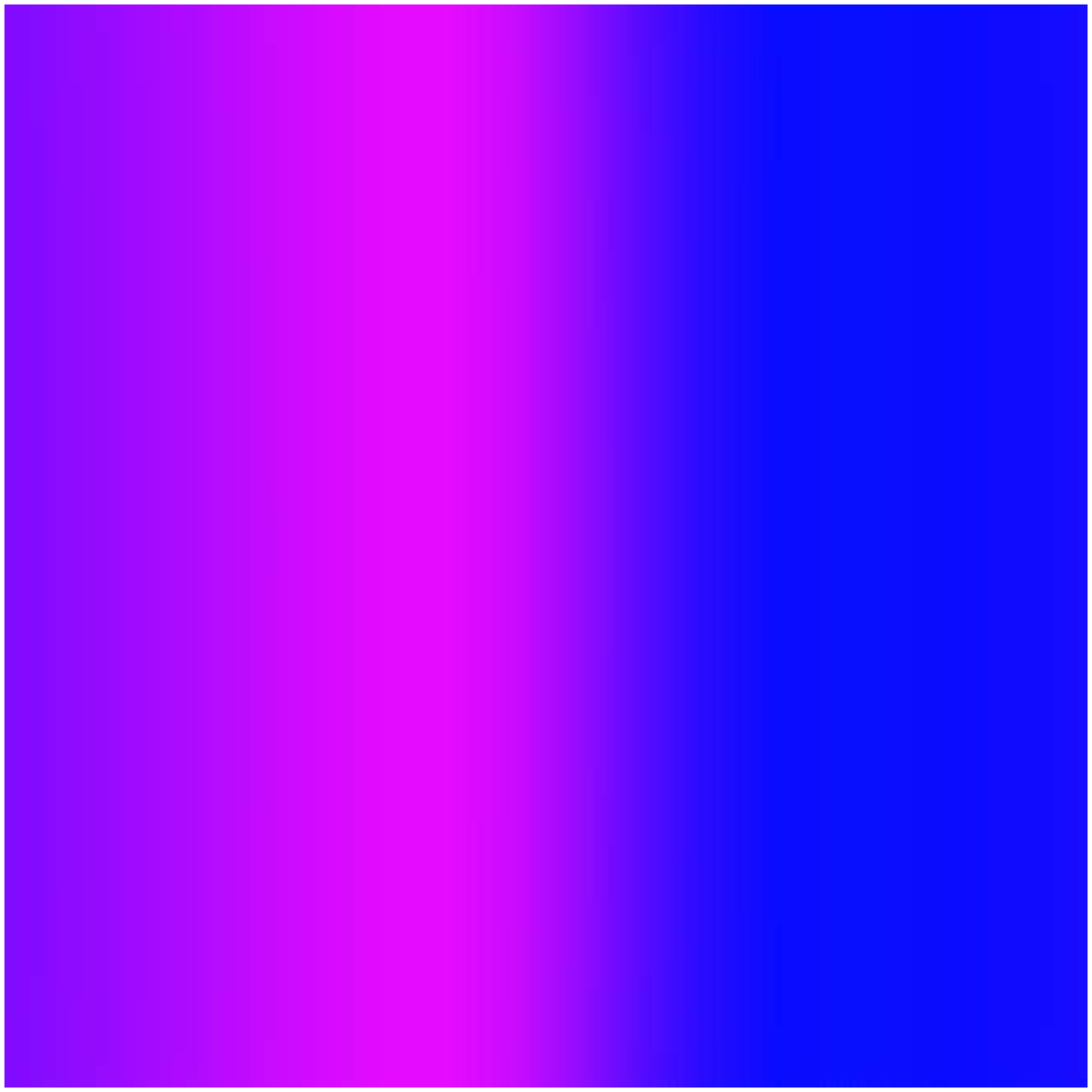}}
\resizebox{7.5cm}{7.5cm}{\includegraphics{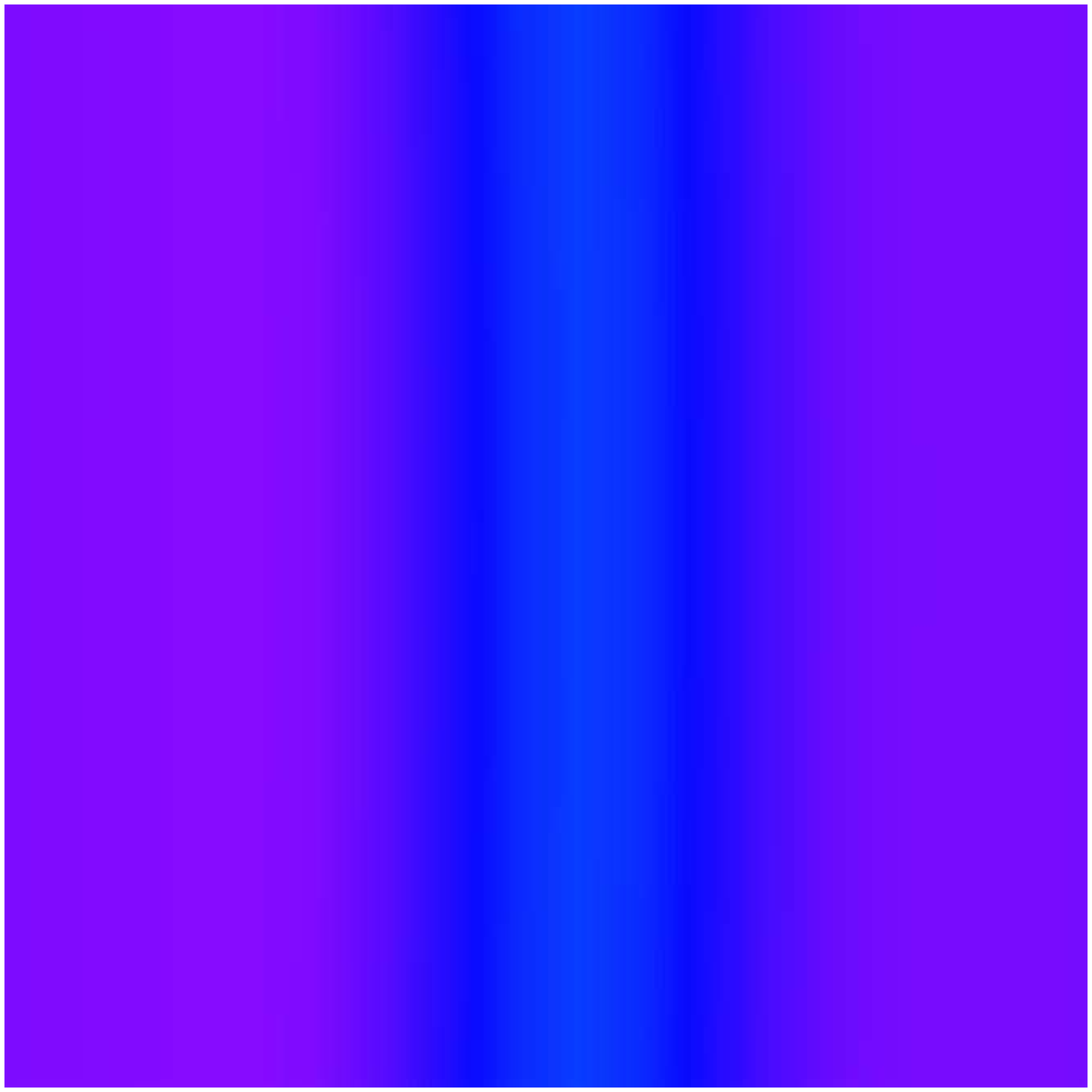}}
\resizebox{\textwidth}{!}{\includegraphics{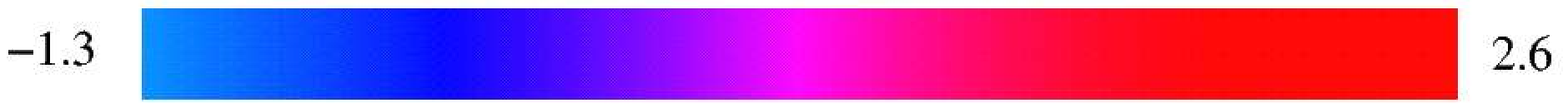}}
\caption{Temperature discontinuity induced by a straight moving string
with an initial velocity $v=0.9$: total (top-left), scalar
(top-right), vector (bottom-left) and tensor (bottom-right).  The
colour scale is given in units of $\Delta = 8\pi
G\mu/c^2$}\label{fig_ks}
\end{figure}

In the realisation shown in figure~\ref{fig_ks}, we minimised these
effects by considering photon propagation in the late Universe with no
cosmological constant,
starting at $\eta = 0.34\eta_0$ and ending at $\eta = 0.41\eta_0$, on
a square patch of sky of $3.2^{\circ}\times3.2^{\circ}$ through a box
of comoving size $L=0.069\eta_0$.  The string was initially moving at a
velocity $v=0.9$, which had redshifted to $v=0.825$ as the photons
crossed the plane of the string. These test simulations included no compensation,
i.e. everything was set to zero initially. 
The temperature jump measured from
the average of the plateaus behind and in front of the string is
approximately
\begin{equation}
\frac{\Delta T}{T} \simeq 53 G\mu/c^2 \, ,
\end{equation}
whereas the theoretical flat-space, vacuum (no matter) prediction for the given
velocities should lie in the range:
\begin{equation}
\frac{\Delta T}{T} \simeq (37-52) G\mu/c^2 \, .
\end{equation}
This is in good agreement with the expanding Universe
simulation, despite the additional physical effects involved.    It appears that the effect
of matter on the scalar and vector modes slightly increases the
anisotropies over the dominant flat-space ISW effect.

\subsection{Loop Anisotropies}

In this subsection, we show the results of small angular scale
simulations of string loops of radius $R=0.5\eta$.
The size of the horizon at the \textit{beginning} of the simulation is
$0.5L$ where $L$ is the size of the box.  The simulations end at
$\eta\approx 10\eta_{dec}$.  Hence, we are dealing with large loops
smaller than the horizon, but initially perturbed by Hubble damping.

The general loop solution in flat-space can be expanded in Fourier
modes, with the gauge conditions yielding constraints on the
relationship between the different modes.  These can be solved if only
the first few harmonics are considered.  The Kibble-Turok
solution~\cite{kibble82} for a loop of length $L$ involves the first
and third harmonics:
\begin{equation}\begin{array}{cl}
\mathbf{x}(\zeta,t) = &
\frac{L}{4\pi}\left\{\hat{\mathbf{e}}_1\left[(1-\kappa)\sin\sigma_- +
\frac{1}{3}\kappa\sin3\sigma_- + \sin\sigma_+\right] \right.\\
&
- \hat{\mathbf{e}}_2\left[(1-\kappa)\cos\sigma_- +
\frac{1}{3}\kappa\cos3\sigma_- + \cos\varphi\cos\sigma_+\right] \\
&
\left. - \hat{\mathbf{e}}_3\left[2\kappa^{1/2}(1-\kappa)^{1/2}\cos\sigma_- +
\sin\varphi\cos\sigma_+\right]\right\} \, ,
\end{array}
\end{equation}
where $\sigma_{\pm} = (2\pi/L)\zeta_{\pm}$ are the left/right moving
modes and the $\hat{\mathbf{e}_i}$ are the cartesian unit vectors.

\begin{figure}
\resizebox{7.5cm}{7.5cm}{\includegraphics{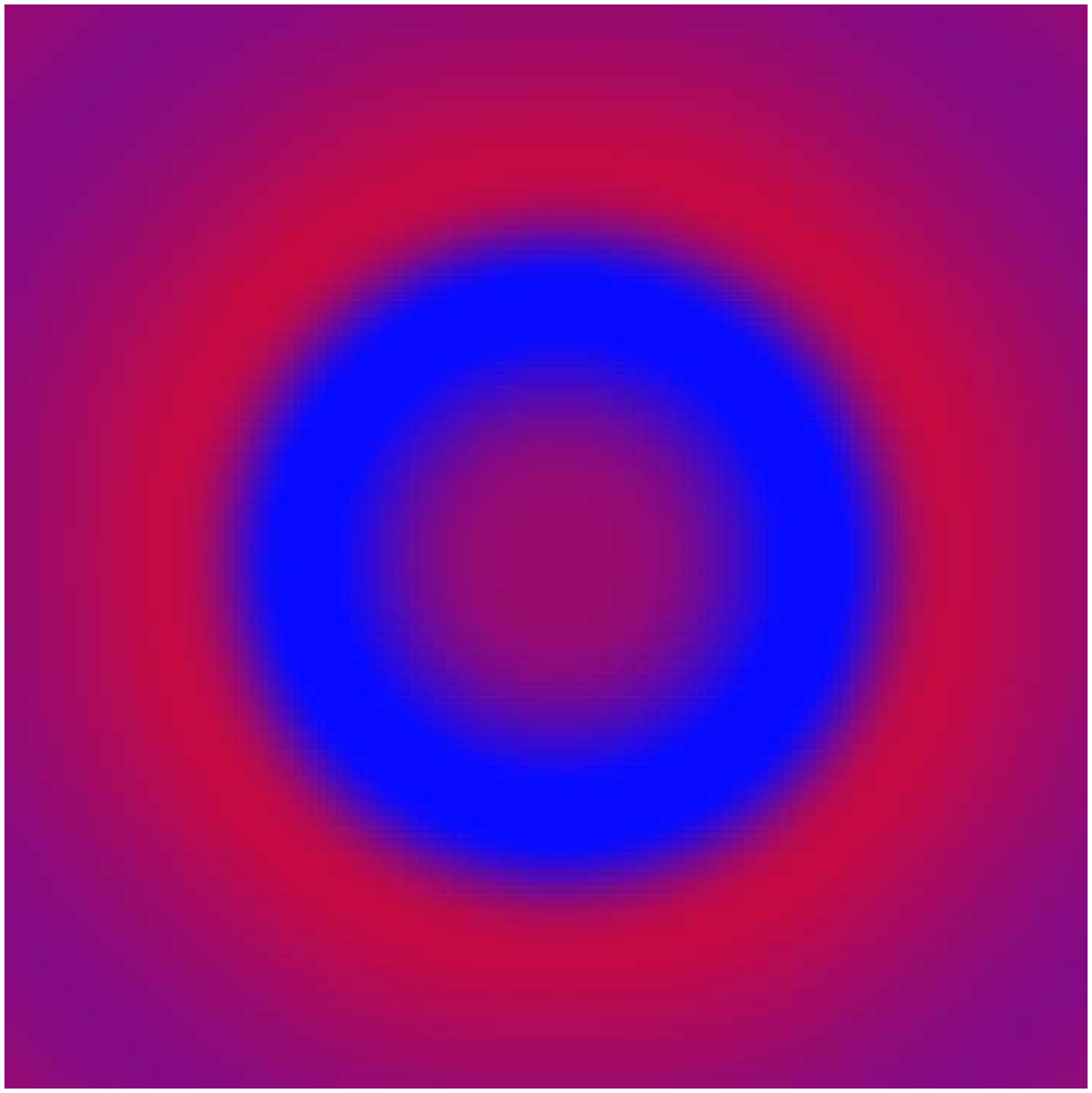}}
\resizebox{7.5cm}{7.5cm}{\includegraphics{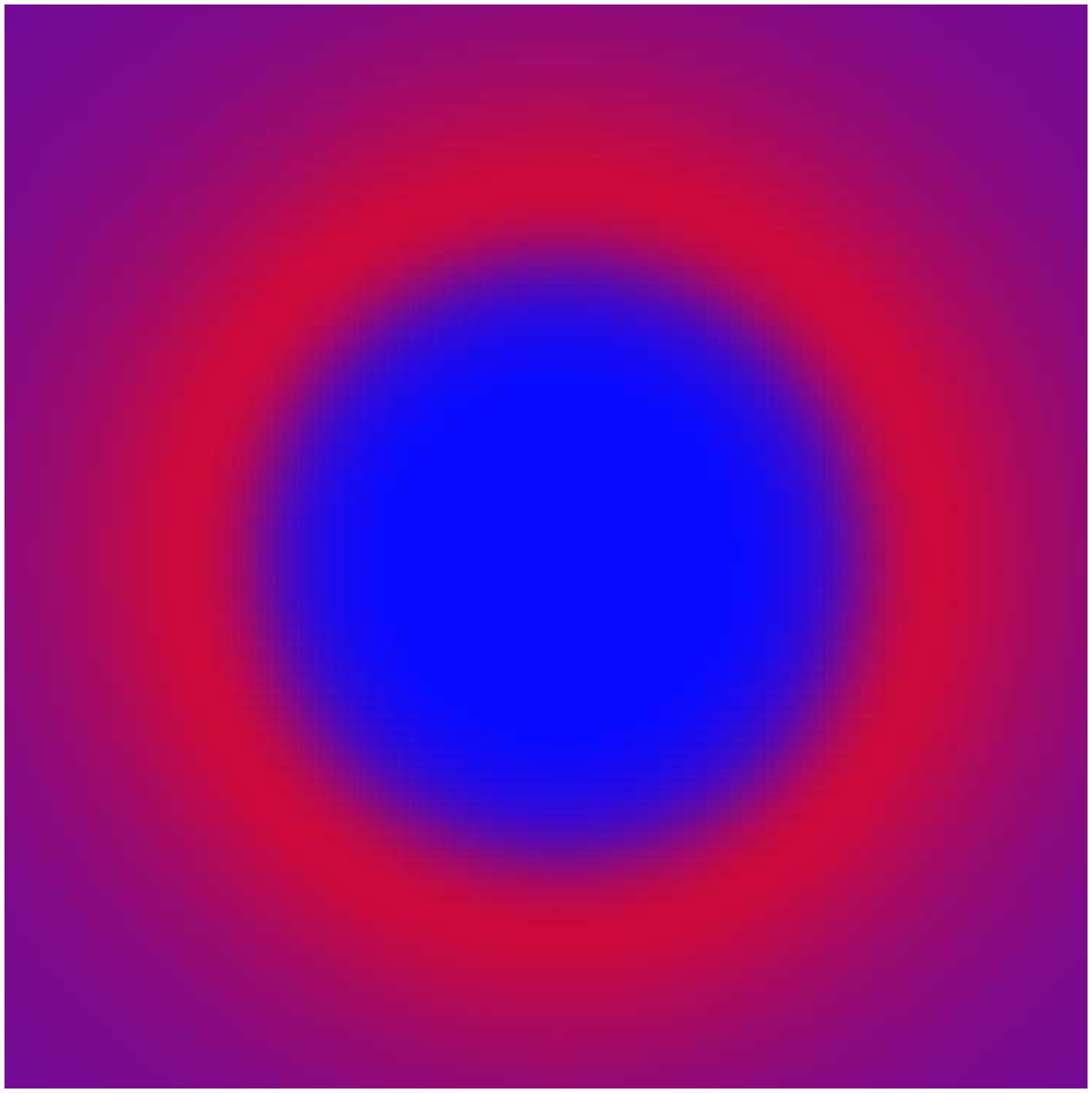}}
\resizebox{7.5cm}{7.5cm}{\includegraphics{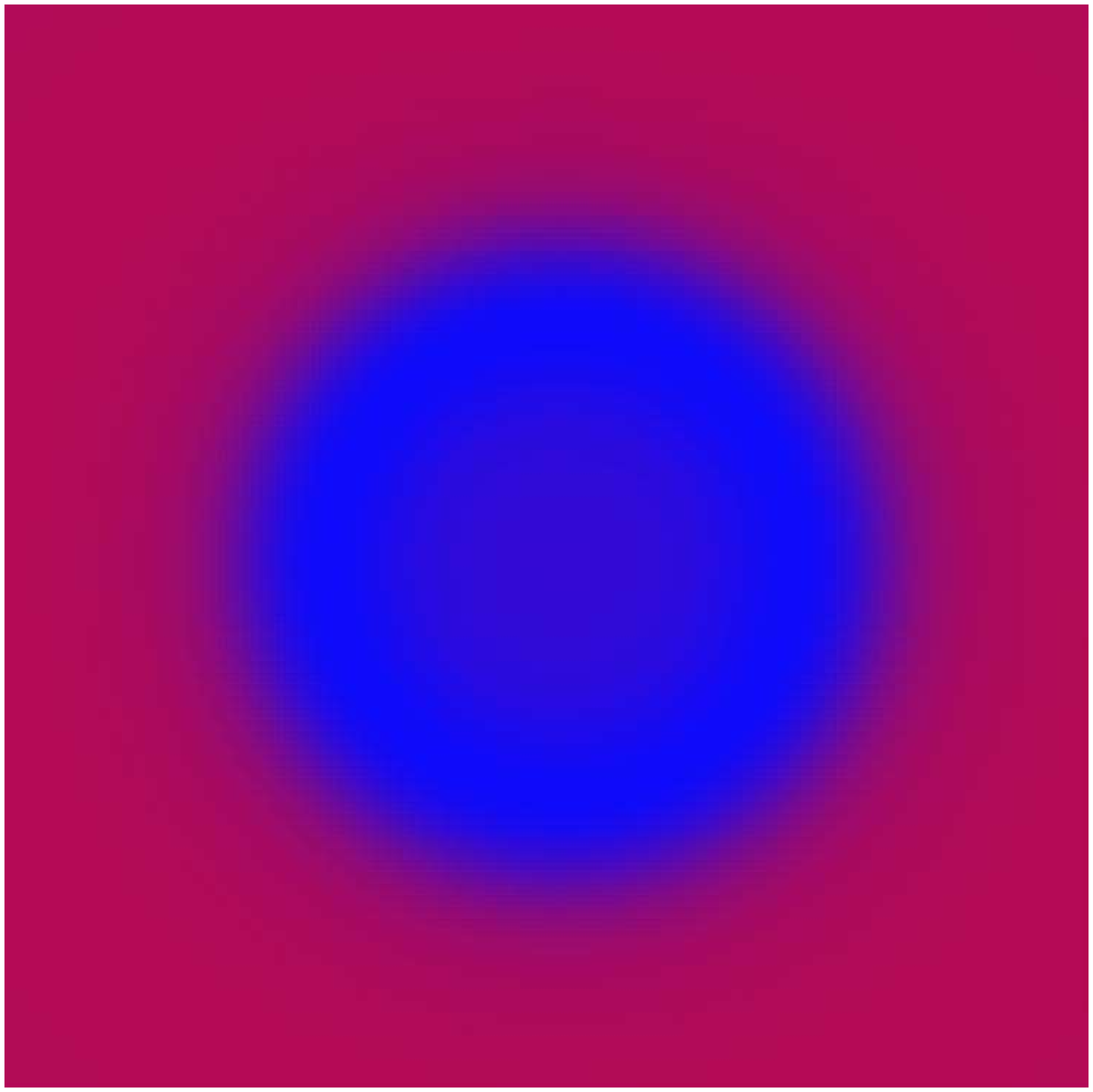}}
\resizebox{7.5cm}{7.5cm}{\includegraphics{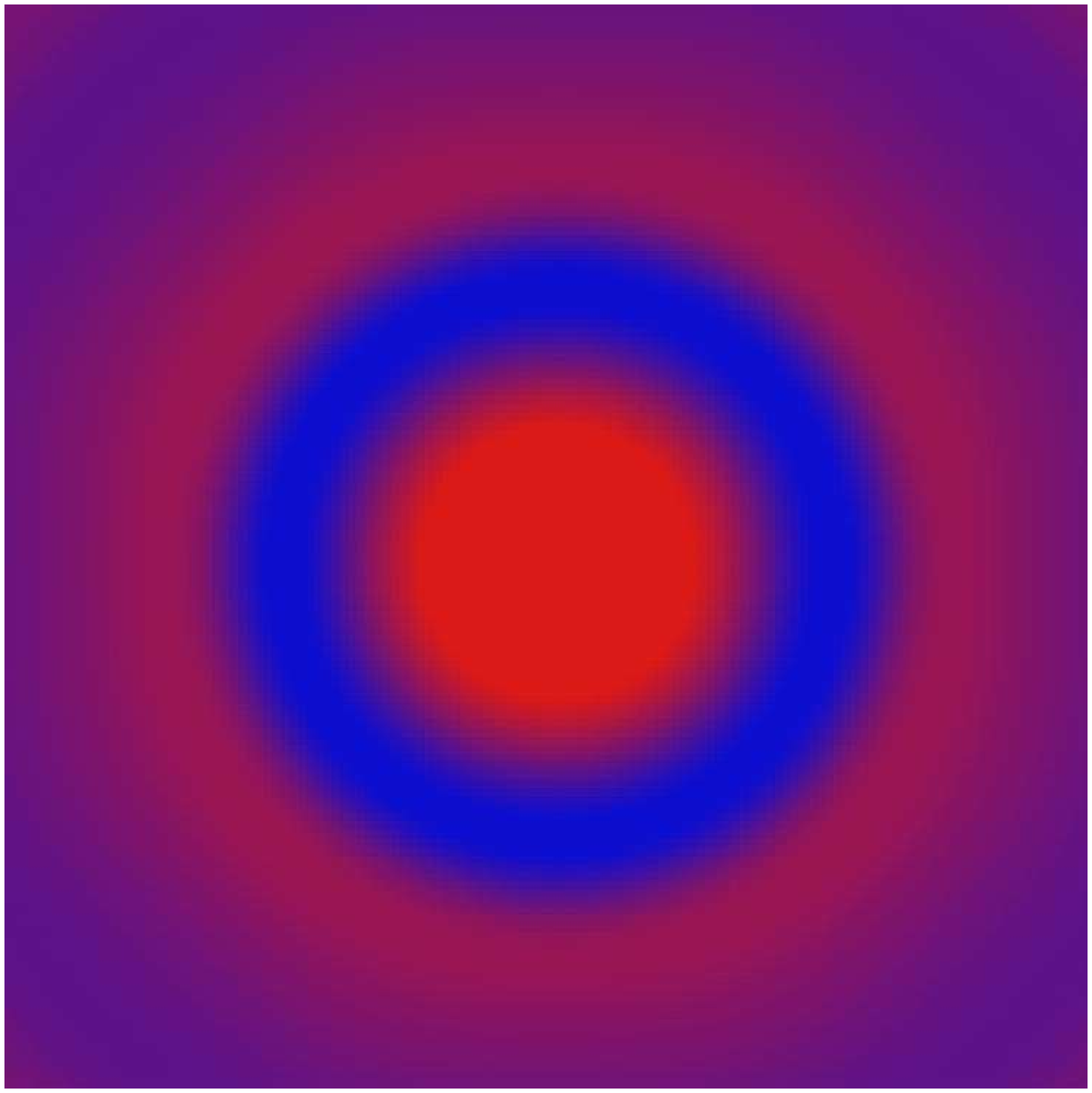}}
\caption{CMB fluctuations seeded by a single loop: total (top-left),
scalar (top-right), vector (bottom-left) and tensor
(bottom-right).}\label{fig_circ}
\end{figure}

Figure~\ref{fig_circ} shows the CMB fluctuations caused by a
perfectly circular loop ($\varphi = \kappa = 0$).
The loop has an initial radius just below the angular scale of the
patch shown in figure~\ref{fig_circ}.  The loop begins stationary and
then accelerates up to a relativistic velocity as it collapses.  At
this point it is crossed by the plane of CMB photons where a clear
Kaiser-Stebbins discontinuity is evident; note that the amplitude of
the temperature increases from large to small radius as the
velocity increases.  There is a cancellation of the non-local scalar,
vector and tensor modes in the interior.  By the time of photon
crossing, causal effects related to the velocity acceleration have not
had time to propagate to the centre.  During one period, the average
velocity of a loop is $<v> = 1/\sqrt{2}$.

\begin{figure}
\resizebox{7.5cm}{7.5cm}{\includegraphics{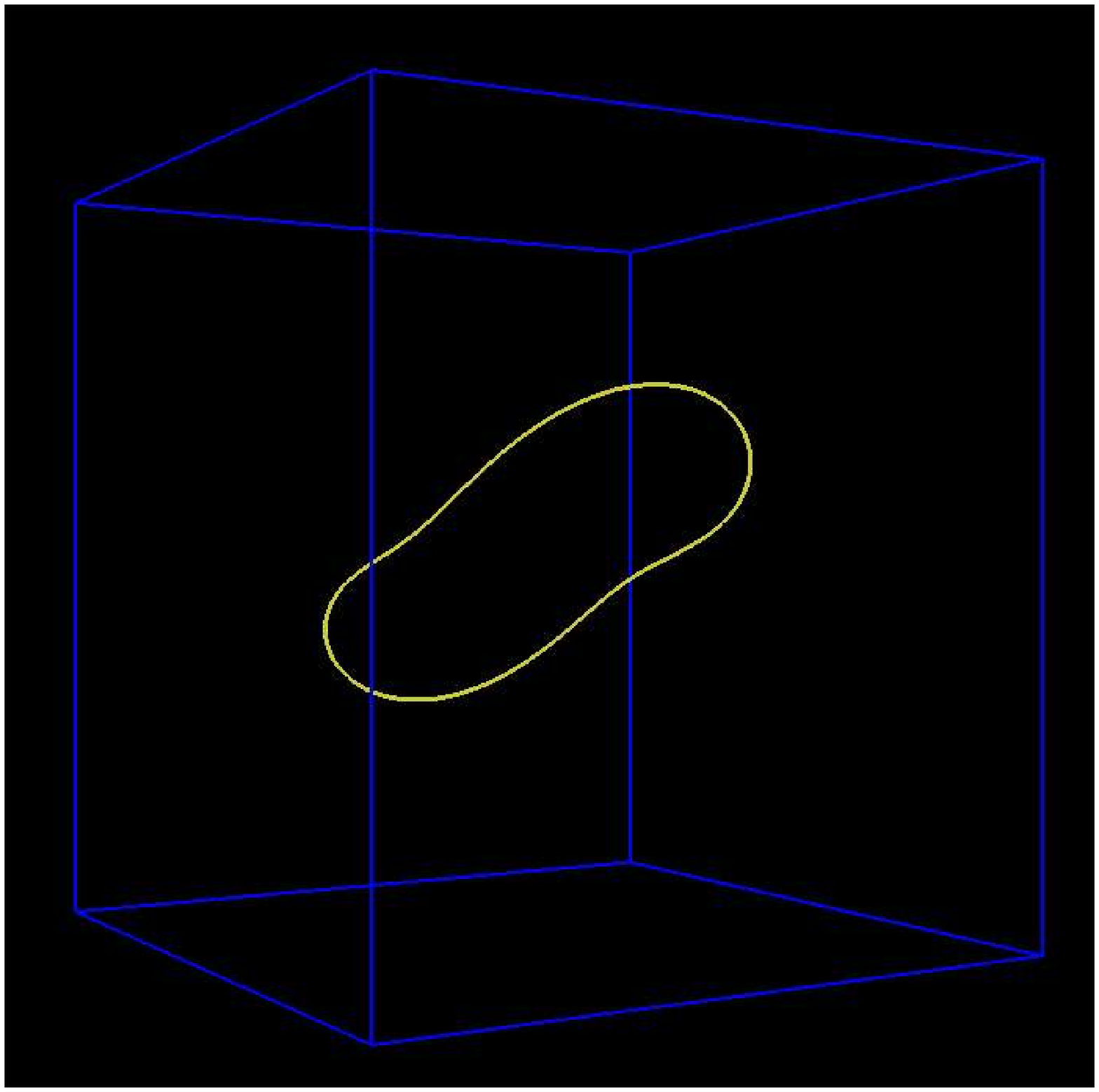}}
\resizebox{7.5cm}{7.5cm}{\includegraphics{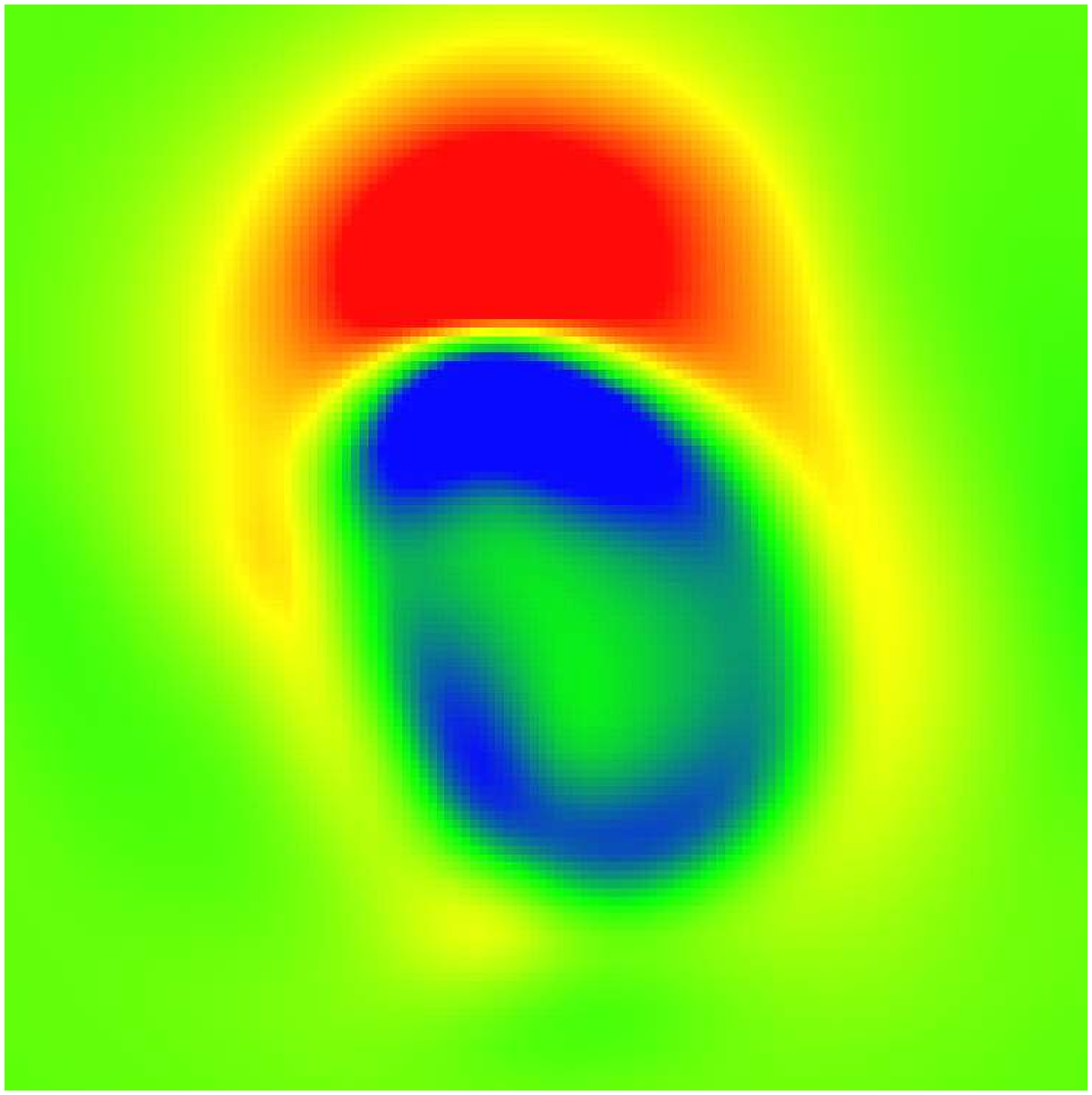}}
\resizebox{7.5cm}{7.5cm}{\includegraphics{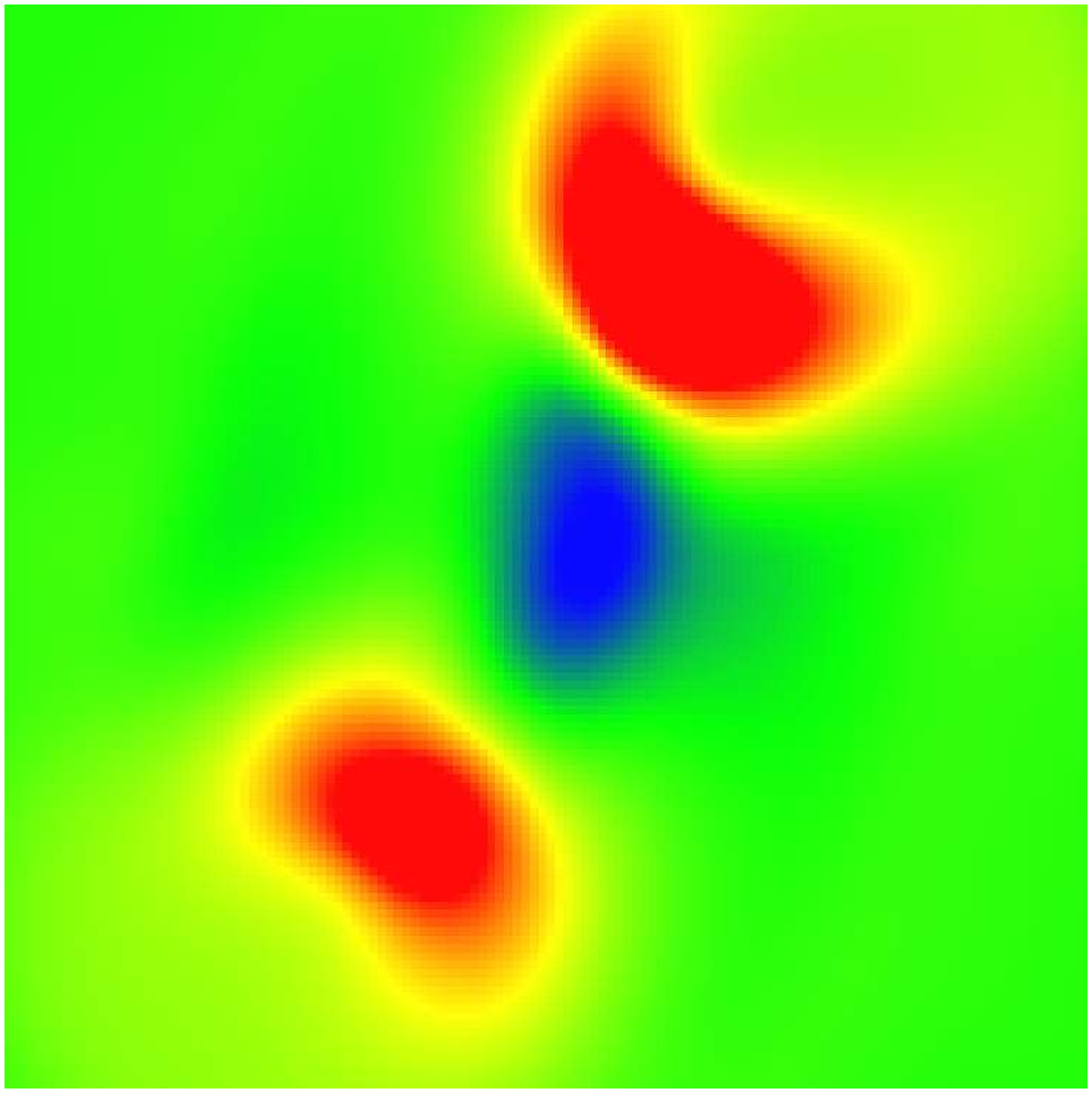}}
\resizebox{7.5cm}{7.5cm}{\includegraphics{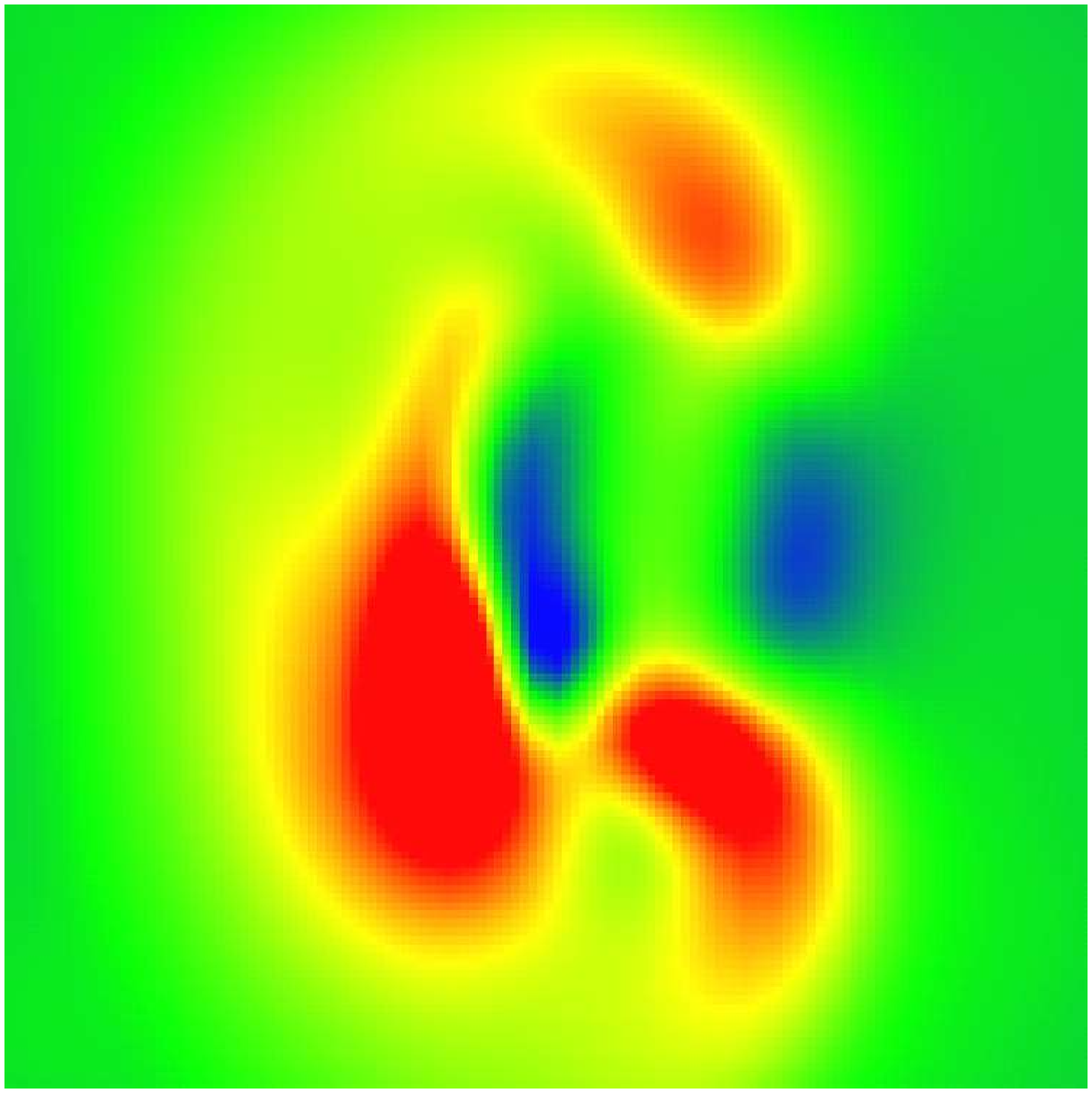}}
\caption{CMB fluctuations seeded by a single (Kibble-Turok) loop (top-left):
front view (top-right), side view (bottom-left) and top view
(bottom-right).}\label{fig_kt}
\end{figure}

Next we consider a loop with $\varphi = \pi/3$ and $\kappa = 1/2$,
which does not possess the simple symmetries of the circular case.
Its initial state is one of maximum radius and minimum velocity from
which it then shrinks, accelerates and begins to rotate and form
cusps.   
Figure~\ref{fig_kt} illustrates the total anisotropy
pattern viewed from three different directions into the box.  In the
top right corner, there are two things to note: when the photons first
cross the loop, there is little effect as it is moving slowly;
whereas there is a strong discontinuity later when the photons pass
the front of the relativistically collapsing loop.  The bottom left
corner shows a side view of the loop, while the bottom right shows a
much more complex anisotropy pattern as cusps form viewed from above.

\section{Discussion}

We have presented the full set of sourced evolution equations with the
Boltzmann hierarchy necessary to study the gravitational effects of
cosmic defects on the CMB.  We have focussed attention on using cosmic
string simulations as the evolving causal source terms in these
equations.  We have developed efficient numerical techniques for the
Green's function computation of high accuracy maps from these string
network simulations.  We have also presented numerical tests of the
full pipeline.  More quantitative results from extensive supercomputer
simulations will be presented in~\cite{lacmb,hrcmb},
featuring large-angle and small-angle maps respectively.

\section*{Acknowledgements}

We are grateful for useful discussions with Gareth Amery, Richard Battye, 
Martin Bucher, Carlos Martins and Proty Wu.
ML acknowledges the
support of the FCAR Fund; the Cambridge Commonwealth Trust; ORS;
Peterhouse; the Canadian Centennial Scholarship Fund; and the
Cambridge Philosophical Society.  This work was supported by PPARC
grant no.  PPA/G/O/1999/00603.  Simulations were performed on COSMOS,
the Origin 3000 supercomputer, owned by the UK cosmology consortium
and funded by SGI/Cray Research, HEFCE and PPARC.

\bibliography{refs,myrefs}

\end{document}